\pdfoutput=1

\documentclass[11pt,twoside,a4paper,cmspaper,final,collab]{cms-tdr}
\def\svnVersion{464702P}\def\cmsCernNoTag{CERN-EP-2017-155}\def\cmsCernDate{\today}\def\cmsMessage{Published in Physics Letters B as \href{http://dx.doi.org/10.1016/j.physletb.2018.04.036}{\doi{10.1016/j.physletb.2018.04.036.}}}

\begin{document}\cmsNoteHeader{B2G-17-007}

\hyphenation{had-ron-i-za-tion}
\hyphenation{cal-or-i-me-ter}
\hyphenation{de-vices}
\RCS$HeadURL: svn+ssh://svn.cern.ch/reps/tdr2/papers/B2G-17-007/trunk/B2G-17-007.tex $
\RCS$Id: B2G-17-007.tex 452451 2018-03-23 21:14:56Z alverson $
\newlength\cmsFigWidth
\ifthenelse{\boolean{cms@external}}{\setlength\cmsFigWidth{0.85\columnwidth}}{\setlength\cmsFigWidth{0.4\textwidth}}
\ifthenelse{\boolean{cms@external}}{\providecommand{\cmsLeft}{top\xspace}}{\providecommand{\cmsLeft}{left\xspace}}
\ifthenelse{\boolean{cms@external}}{\providecommand{\cmsRight}{bottom\xspace}}{\providecommand{\cmsRight}{right\xspace}}
\providecommand{\NA}{\ensuremath{\text{---}}}
\newcommand{\x}{\ensuremath{\phantom{0}}}
\newcommand{\y}{\ensuremath{\phantom{.}}}

\cmsNoteHeader{B2G-17-007}
\title{Search for single production of a vector-like T quark decaying to a Z boson and a top quark in proton-proton collisions at $\sqrt{s} = 13\TeV$}

\date{\today}

\abstract{
A search is presented for single production of a vector-like quark (T) decaying to a Z boson and a top quark, with the Z boson decaying leptonically and the top quark decaying hadronically. The search uses data collected by the CMS experiment in proton-proton collisions at a center-of-mass energy of 13\TeV in 2016, corresponding to an integrated luminosity of 35.9\fbinv. The presence of forward jets is a particular characteristic of single production of vector-like quarks that is used in the analysis. For the first time, different T quark width hypotheses are studied, from negligibly small to 30\% of the new particle mass. At the 95\% confidence level, the product of cross section and branching fraction is excluded above values in the range 0.26--0.04\unit{pb} for T quark masses in the range 0.7--1.7\TeV, assuming a negligible width. A similar sensitivity is observed for widths of up to 30\% of the T quark mass. The production of a heavy \PZpr boson decaying to Tt, with $\mathrm{T}\to\PQt\Z$, is also searched for, and limits on the product of cross section and branching fractions for this process are set between 0.13 and 0.06\unit{pb} for \PZpr boson masses in the range from 1.5 to 2.5\TeV.
}

\hypersetup{%
pdfauthor={CMS Collaboration},%
pdftitle={Search for single production of vector-like quarks decaying to a Z boson and a top or a bottom quark in proton-proton collisions at 13 TeV},%
pdfsubject={CMS},%
pdfkeywords={CMS, physics, vector-like quarks}}
\maketitle
\section{Introduction}
A possible extension of the standard model (SM), able to address some of the problems related to the nature of electroweak symmetry breaking, involves heavy particles called vector-like quarks (VLQs)~\cite{Aguilar-Saavedra:2013qpa, AguilarSaavedra:2009es, DeSimone:2012fs, Matsedonskyi:2014mna, Buchkremer:2013bha}. Unlike the chiral fermions of the SM, these new particles do not obtain mass through a Yukawa coupling but through a direct mass term of the form $m\overline{\psi}\psi$. This means that they are not excluded by precision SM measurements as are fourth-generation chiral quarks~\cite{Eberhardt:2012sb}.

Previous searches for VLQs have been performed by both the ATLAS~\cite{Aad:2015kqa, Aad:2015mba, Aad:2014efa, Aad:2016shx, Aad:2016qpo, Aad:2015voa, Aaboud:2017qpr, Aaboud:2017zfn} and CMS~\cite{Khachatryan:2015oba, Khachatryan:2015gza, Chatrchyan:2013wfa, Sirunyan:2017ezy, Khachatryan:2016vph, Sirunyan:2016ipo, Sirunyan:2017tfc, Sirunyan:2017usq} Collaborations, as well as by the D0~\cite{Abazov:2010ku, Abazov:2011vy} and CDF~\cite{PhysRevD.76.072006, PhysRevLett.104.091801, Aaltonen:2011na, Aaltonen:2011rr, Aaltonen:2011vr, Aaltonen:2011tq} Collaborations.

\begin{figure}[!b]
\centering
\includegraphics[scale=0.40]{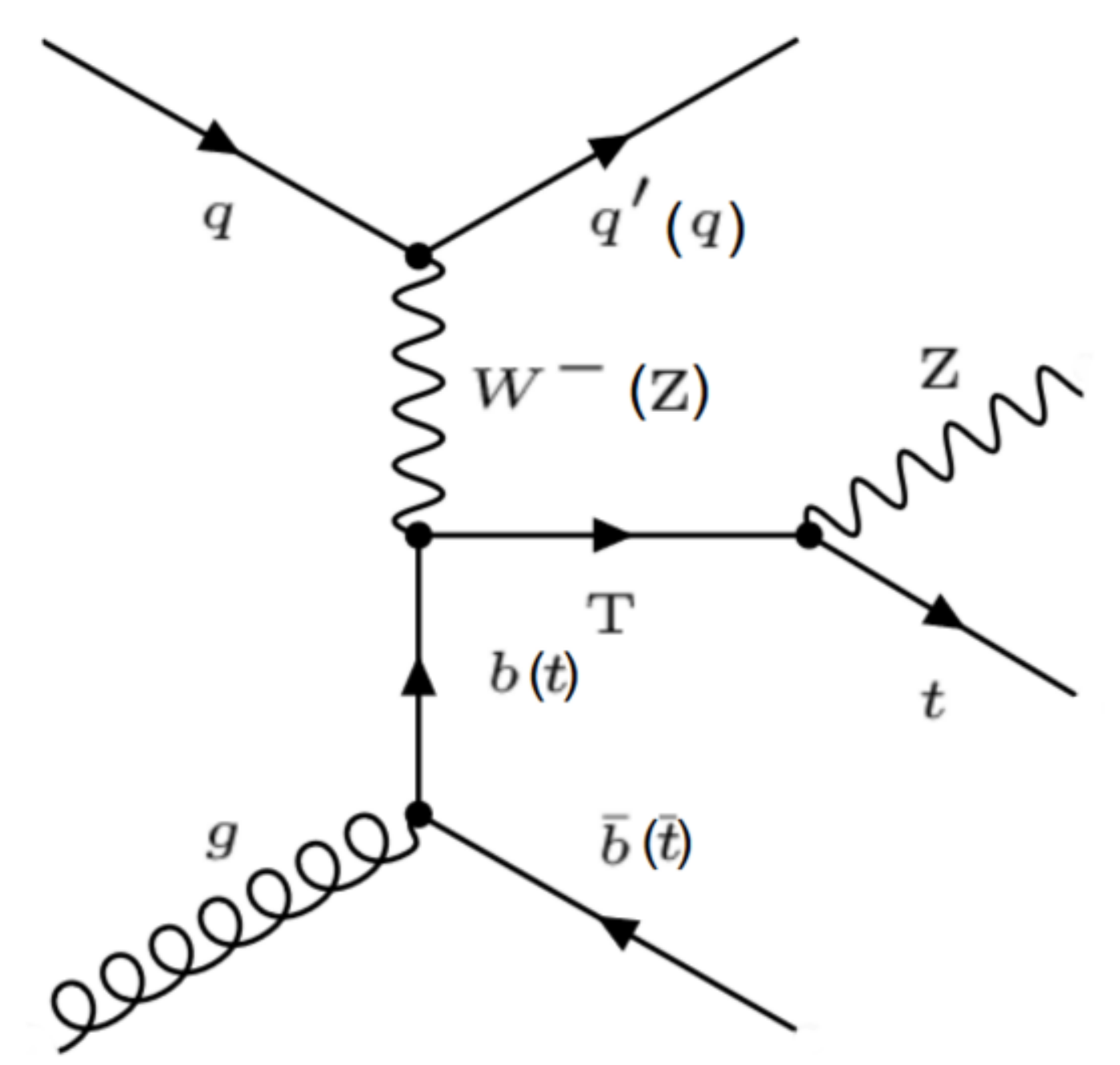}
\ifthenelse{\boolean{cms@external}}{\\}{\hspace{15mm}}
\includegraphics[scale=0.40]{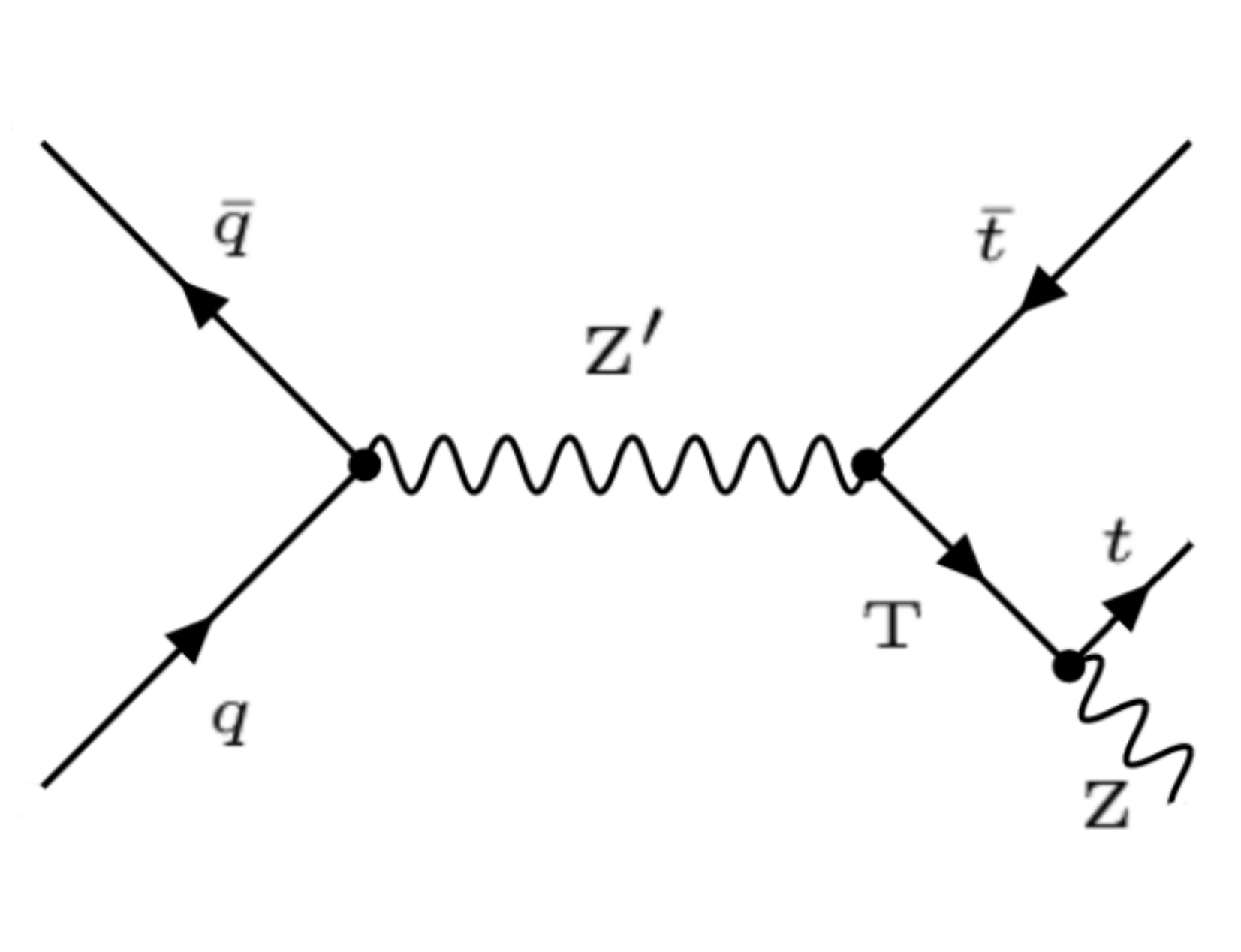}
\caption{
Leading-order Feynman diagrams for the production of a single vector-like T quark and its decay to a Z boson and a t quark, either in association with a b quark or a t quark (\cmsLeft), or in the decay of a \PZpr boson to Tt (\cmsRight).}\label{fig:feynman}
\end{figure}

We study the single production of vector-like T quarks with charge $+2/3$ that decay to a Z boson and a t quark. We search for a final state with a Z boson decaying to electrons or muons, and a t quark producing jets via the decay $ \PQt \to \PW\PQb \to \PQq^{\prime}\PAQq\PQb$. An example of a leading-order (LO) Feynman diagram for the single production of a T quark in association with either a b quark, denoted T(b), or a t quark, denoted T(t), is shown in Fig.~\ref{fig:feynman} (\cmsLeft). The three decay channels of the T quark into SM particles are bW, tZ, and tH. If the T is a singlet of the SM, the equivalence theorem~\cite{Lee:1977eg} implies that the branching fractions for the three decay modes of the T quark are approximately 0.5, 0.25, and 0.25, respectively. If the T is a doublet of the SM, the decay modes are tZ and tH, each with a branching fraction of 0.5.

The T quark could be singly produced in association with either a t or a b quark and an additional quark would be produced in the forward region of the detector. The coupling coefficients of the T quark to SM particles are denoted C(bW) for the T(b) process, and C(tZ) for the T(t) process. The production cross section of the T quark depends on its mass and width, as well as on these couplings. The T quark can have both left-handed (LH) and right-handed (RH) couplings to SM particles. In the case of a singlet T quark, the RH chirality is suppressed by a factor proportional to the SM quark mass divided by the T quark mass. In the case of a doublet T quark, it is the LH chirality that is suppressed~\cite{Cacciapaglia:2011fx}.

The present search is also sensitive to the production of a T quark together with a t quark in the decay of a heavy neutral spin-1 \PZpr boson~\cite{Bini:2011zb, Accomando:2010fz, Greco:2014aza}. A LO Feynman diagram for this production mode is shown in Fig.~\ref{fig:feynman} (\cmsRight). This channel was also considered in Refs.~\cite{Sirunyan:2017ezy, Sirunyan:2017bfa}.

This search follows a strategy similar to that used by Ref.~\cite{Sirunyan:2017ezy}. However, significant improvements to the sensitivity of the method have been made by employing a categorization based on the presence of forward jets, and by analyzing the mass spectrum of reconstructed T quark candidates, $m_{\text{tZ}}$, in events where the t quark products are highly Lorentz-boosted and therefore are reconstructed as a single, large-radius jet. The present analysis also benefits from the much larger data sample recorded in 2016. This paper also includes the first results assuming a T quark with a nonnegligible decay width that varies between 10 and 30\% of the T quark mass.

\section{The CMS detector, data, and simulation}\label{sec:CMS}

The general-purpose CMS detector operates at one of the four interaction points of the LHC. Its central feature is a 3.8\unit{T} superconducting solenoid magnet with an inner diameter of 6\unit{m}. The following subdetectors are found within the magnet volume: a silicon tracker, a crystal electromagnetic calorimeter (ECAL), and a brass and scintillator hadron calorimeter (HCAL). Muons are measured in gas-ionization detectors embedded in the steel flux-return yoke outside the solenoid. In addition, the CMS detector has extensive forward calorimetry: two steel and quartz-fiber hadron forward calorimeters that extend the HCAL coverage to regions close to the beam pipe, and cover the pseudorapidity range $3.0 < \abs{\eta} < 5.2$. A more detailed description of the CMS detector, together with a definition of the coordinate system and kinematic variables, can be found in Ref.~\cite{Chatrchyan:2008aa}.

This analysis is based on the data collected by the CMS experiment in proton-proton collisions at a center-of-mass energy of 13\TeV in 2016, corresponding to an integrated luminosity of 35.9\fbinv. Events with a Z boson decaying to muons are selected online by requiring the presence of an isolated muon with transverse momentum $ \pt > 24\GeV$. Events with the Z boson decaying to electrons are selected online if an electron is reconstructed with $\pt > 115\GeV$. It is possible to use this relatively high \pt threshold without losing signal efficiency, since the electrons of interest arise from the decay of a heavy resonance.

Background events are generated using the next-to-LO (NLO) generator \MGvATNLO 2.2.2~\cite{Alwall:2014hca} for Z/$\gamma$*+jets, \ttbar+V, and tZq processes,
and the NLO generator \POWHEG~2.0~\cite{Nason:2004rx, Frixione:2007vw, Alioli:2010xd, Alioli:2011as} for \ttbar and single t quark production. They are interfaced with \PYTHIA~8.212~\cite{Sjostrand:2014zea}, with the tune \textsc{cuetp8m2t4}~\cite{CMS-PAS-TOP-16-021} used for the description of parton hadronization and fragmentation. Events for diboson production are generated at LO using \MADGRAPH~5.2 and at NLO with \POWHEG~2.0. Simulated events are normalized to NLO cross sections for all processes except for \ttbar, single t quark production and diboson (WW only) processes, where next-to-NLO values are used.

Signal events with the T quark produced either directly or in the decay of a \PZpr boson are generated at LO using \MADGRAPH~5.2 interfaced to \PYTHIA~8.212. For the single production of the T quark, different T quark width hypotheses are considered: negligibly small and larger widths (10, 20, and 30\% of the T quark mass). Spin correlations are treated in the decay with {\sc madspin}~\cite{Artoisenet:2012st}.

In the case where the T and \PZpr particles are generated with narrow widths, i.e., negligibly small with respect to the experimental reconstructed mass resolution, T quark masses $m_{\mathrm{T}}$ between 0.7 and 1.7\TeV in steps of 0.1\TeV, and \PZpr masses $m_{\text{Z}^{\prime}}$ of 1.5, 2.0, and 2.5\TeV are considered. The singlet T(b) signal process with LH couplings to SM particles, and doublet T(t) signal process with RH couplings, are generated. Theoretical cross sections for the narrow-width T quark assumption are listed in Table~\ref{tab:CrossSec}, calculated following the procedures described in Ref.~\cite{Matsedonskyi:2014mna}, where a simplified approach is used to provide a model-independent interpretation of experimental results. The width of the VLQ is negligible compared to the experimental mass resolution for C(bW) and C(tZ) couplings ${\leq}$0.5.

Signals for T quarks with larger widths (10, 20, and 30\% of the T quark mass) are generated in the same mass range but in steps of 0.2\TeV. The effect of the finite-width approximation is evaluated using a modified version of the model constructed by the authors of Refs.~\cite{Buchkremer:2013bha,Fuks:2016ftf,Oliveira:2014kla}. Modifications of the published versions were necessary to provide a simulation of the full 2~$\to$~4 process, i.e., $\Pp\Pp \to \mathrm{T}\PQb\PQq/\mathrm{T}\PQt\PQq \to \PQt\PZ\PQb\PQq/\PQt\PZ\PQt\PQq$, in the finite-width hypothesis. It has been verified that the interference of the 2~$\to$~4 process with the SM background processes is negligible.

\begin{table}[!t]
\centering
\topcaption{Theoretical cross sections at next-to-leading order for single production of a T quark in association with a b or t quark for the benchmark masses considered in the analysis, with the couplings set to 0.5 and using the narrow-width T quark assumption, calculated following the procedures described in Ref.~\cite{Matsedonskyi:2014mna}. The cross sections do not depend on the chirality of the T quark. The narrow-width assumption is valid for any value of the couplings less than or equal to 0.5.}\label{tab:CrossSec}
\ifthenelse{\boolean{cms@external}}{\resizebox{\columnwidth}{!}}{}
{
\begin{tabular}{ c c c }
\hline
$m_{\mathrm{T}}$ [\TeVns{}] & $\sigma$(pp~$\to$~Tbq~$\to$~tZbq) [pb] & $\sigma$($\Pp\Pp\to\mathrm{T}\PQt\PQq$~$\to$~tZtq) [pb] \\
\hline
0.7 & 0.364 & 0.063 \\
0.8 & 0.241 & 0.046 \\
0.9 & 0.170 & 0.034 \\
1.0 & 0.122 & 0.026 \\
1.1 & 0.085 & 0.019 \\
1.2 & 0.062 & 0.015 \\
1.3 & 0.045 & 0.011 \\
1.4 & 0.034 & 0.009 \\
1.5 & 0.026 & 0.007 \\
1.6 & 0.019 & 0.006 \\
1.7 & 0.015 & 0.004 \\
\hline
\end{tabular}
}

\end{table}

In the general case, the total production cross section for a T quark with a finite width (FW) can be written as:
\begin{equation}\label{equation}
\ifthenelse{\boolean{cms@external}}
{
\begin{split}
\sigma_{\text{FW}}(\text{C}_1,\text{C}_2,m_{\mathrm{T}},\Gamma(\text{C}_1,\text{C}_2,\text{C}_i,m_{\mathrm{T}},m_j)) = \ \ \ \ \ \ \ \ \ \ \\ \text{C}_1^2~\text{C}_2^2~\tilde\sigma_{\text{FW}}(m_{\mathrm{T}},\Gamma(\text{C}_1,\text{C}_2,\text{C}_i,m_{\mathrm{T}},m_j))\,,
\end{split}
}
{
\sigma_{\text{FW}}(\text{C}_1,\text{C}_2,m_{\mathrm{T}},\Gamma(\text{C}_1,\text{C}_2,\text{C}_i,m_{\mathrm{T}},m_j))=\text{C}_1^2~\text{C}_2^2~\tilde\sigma_{\text{FW}}(m_{\mathrm{T}},\Gamma(\text{C}_1,\text{C}_2,\text{C}_i,m_{\mathrm{T}},m_j))\,,
}
\end{equation}
where $\Gamma(\text{C}_1,\text{C}_2,\text{C}_i,m_{\mathrm{T}},m_j)$ is the width of the T quark, $\text{C}_1$ and $\text{C}_2$ are its couplings to SM quarks and bosons in the specific single-production process under consideration, $\text{C}_i$ summarizes other possible couplings that allow the T to decay to other final states, and the quantities $m_j$ represent the masses of the decay products of the T quark. The $\tilde\sigma_{\text{FW}}$ is the ``reduced cross section'' and it corresponds to the physical cross section after factorizing the production cross section and the decay couplings. For the process $\Pp\Pp \to \mathrm{T}\PQt\PQq \to \PQt\PZ\PQt\PQq $ the couplings are $\text{C}_1 = \text{C}_2 = (\text{g}_{\text{w}}/2) \, \text{C(tZ)}$, while for $\Pp\Pp\to \mathrm{T}\PQb\PQq \to \PQt\PZ\PQb\PQq$ the couplings are $\text{C}_1 = (\text{g}_{\text{w}}/2) \, \text{C(bW)}$ and $\text{C}_2 = (\text{g}_{\text{w}}/2) \,  \text{C(tZ)}$. The normalisation factor $\text{g}_{\text{w}}/2$ has been introduced to properly compare the couplings as defined in Ref.~\cite{Matsedonskyi:2014mna} and in Eq.~(\ref{equation}). In Table~\ref{tab:CrossSec2}, the values for $\tilde\sigma_{\text{FW}}$ are shown together with the cross sections for the singlet T(b) and doublet T(t) signals used to interpret the results. These cross sections are calculated by fixing the branching fractions of the T to the expected values in the narrow-width approximation, as described above and in Ref.~\cite{Matsedonskyi:2014mna}. This choice corresponds to different sets of couplings than the ones used in the narrow width approximation.

\begin{table*}[!h]
\centering
\topcaption{Theoretical reduced cross sections $\tilde\sigma_{\text{FW}}$ for single production of a T quark with a b or a t quark, where the T quark decays to tZ and its width is 10, 20, and 30\% of its mass, for the benchmark masses considered in the analysis. The corresponding leading order cross sections $\sigma$ for the specified production and decay are shown in parentheses.}\label{tab:CrossSec2}
\begin{tabular}{ c c c c c c c }
\hline
\multirow{2}[0]{*}{$m_{\mathrm{T}}$ [\TeVns{}]} & \multicolumn{3}{c}{$~\tilde\sigma_{\text{FW}}$ ($\sigma$) for $\Pp\Pp\to\mathrm{T}\PQb\PQq\to\PQt\Z\PQb\PQq$ [pb]}  & \multicolumn{3}{c}{$~\tilde\sigma_{\text{FW}}$ ($\sigma$) for $\Pp\Pp\to\mathrm{T}\PQt\PQq\to\PQt\Z\PQt\PQq$ [pb]} \\
 & 10\%  & 20\%  & 30\%   & 10\%  & 20\%  & 30\%  \\\hline
0.8  & 226 (0.675) & 108 (0.650) & 70 (0.631)  & 19  (0.144) & 9.3 (0.139) & 6.0 (0.135) \\
1.0  & 183 (0.314) &  \x87 (0.299) & 55 (0.284)  & 17  (0.075) & 7.9 (0.072) & 5.0 (0.069) \\
1.2  & 145 (0.158) &  \x68 (0.149) & 43 (0.141)  & 14  (0.042) & 6.4 (0.039) & 4.1 (0.037) \\
1.4  & 112 (0.084) &  \x52 (0.079) & 33 (0.074)  & 11  (0.024) & 5.0 (0.022) & 3.2 (0.021) \\
1.6  &  \x85 (0.047) &  \x39 (0.043) & 29 (0.041)  & 8.2 (0.014) & 3.8 (0.013) & 2.4 (0.012) \\
\hline
\end{tabular}
\end{table*}

The generated events are passed through a simulation of the CMS detector based on \GEANTfour~\cite{Agostinelli:2002hh, Allison:2006ve}. The number of additional interactions in the same or adjacent bunch crossings (pileup) is included in simulation with a distribution of the number of additional interactions matching that observed in data. Samples are generated using the {\sc nnpdf}~3.0~\cite{Ball:2014uwa} parton distribution function (PDF) sets, matching the perturbative order used in simulation.

\section{Object reconstruction}\label{sec:object}
Primary vertices are reconstructed using a deterministic annealing filter algorithm~\cite{Chatrchyan:2014fea}. The reconstructed vertex with the largest value of summed physics-object $\pt^2$ is taken to be the primary $\Pp\Pp$ interaction vertex. The physics objects are the objects returned by a jet finding algorithm~\cite{Cacciari:2008gp,Cacciari:2011ma} applied to all charged tracks associated with the vertex, plus the corresponding associated missing transverse momentum. Selected events are required to have this primary vertex within 24\unit{cm} of the center of the detector along the $z$-direction, and within 2\unit{cm} in the $x$-$y$ plane.

A particle-flow (PF) algorithm~\cite{Sirunyan:2017ulk} is used to identify and to reconstruct charged and neutral hadrons, photons, muons, and electrons, through an optimal combination of the information from the entire detector.

Electron candidates are reconstructed by combining the information from the ECAL and from the silicon tracker~\cite{Khachatryan:2015hwa}. Electrons are then selected if they are isolated and if they have $ \pt > 20\GeV$ and pseudorapidity $\abs{\eta} < 2.5$. Additional requirements are applied to the energy distribution in the ECAL, to the geometrical matching of the tracker information to the ECAL energy cluster, on the impact parameters of the charged tracks, and on the ratio of the energies measured in the HCAL and the ECAL in the region around the electron candidate. The leading electron is required to have $\pt > 120\GeV$, in order to be in the region where the trigger is close to 100\% efficiency.

Muon candidates are reconstructed by combining in a global fit the information from the silicon tracker and the muon system~\cite{Chatrchyan:2012xi}. Muons are then required to be isolated, to satisfy $ \pt > 20\GeV$ and $\abs{\eta} < 2.4$, and to pass additional identification criteria based on the track impact parameter, the quality of the track reconstruction, and the number of hits recorded in the tracker and the muon systems. Like the leading electron, the leading muon is required to have $ \pt > 120\GeV$.

For both muons and electrons, a lepton isolation variable is used to reduce background from events in which a jet is misidentified as a lepton. This variable is defined as the scalar sum of the \pt of the charged and neutral hadrons and photons in a cone of size $\Delta R = \sqrt{\smash[b]{(\Delta\eta)^2+(\Delta\phi)^2}}$ around the original lepton track, corrected for the effects of pileup~\cite{Chatrchyan:2012xi, Khachatryan:2015hwa}, and divided by the lepton \pt. The cone size is 0.4 for muons and 0.3 for electrons.

Jet candidates are clustered from the PF candidates using the anti-\kt clustering algorithm~\cite{Cacciari:2008gp} with distance parameters of 0.4 (``AK4 jets'') and 0.8 (``AK8 jets''). The jet energy scale (JES) is calibrated through correction factors dependent on the \pt, $\eta$, energy density, and area of the jet. The jet energy resolution (JER) for the simulated jets is degraded to reproduce the resolution observed in data. The AK4 jet candidates are required to have $ \pt > 20\GeV$, $\abs{\eta} < 2.4$ and to be separated by $\Delta R > 0.4$ from an identified lepton. The AK8 jet candidates are required to have $ \pt > 180\GeV$, $\abs{\eta} < 2.4$ and to be separated by $\Delta R > 0.8$ from an identified lepton. The AK8 jets may be tagged as coming from a W boson decaying to $\PQq^{\prime}\PAQq$ (denoted ``W jets'') or from a t quark decaying fully hadronically (``t jets''). For the W jets, a pruning algorithm~\cite{Ellis:2009su} is applied. The mass of the jet, after the pruning is performed, is used as a discriminant to select W bosons and reject quark and gluon jets. The discrimination between W jets and jets from quarks and gluons is further improved by requiring the $N$-subjettiness ratio $\tau_{21}$ to be less than 0.6, where  $\tau_{21} = \tau_2 / \tau_1$~\cite{Khachatryan:2014vla}, and the mass of the pruned AK8 jet to be within the range 65--105\GeV. In a similar way, AK8 jets may be identified as arising from the all-jets final state of a t quark. These t jets are required to have $ \pt > 400\GeV$, mass of the jet reconstructed through the modified mass drop tagger algorithm~\cite{Dasgupta:2013ihk, Larkoski:2014wba} between 105 and 220\GeV, and $\tau_{32} = \tau_3 / \tau_2$ less than 0.81. Finally, AK4 jets may be tagged as arising from a b quark (``b jets'') using the combined secondary vertex algorithm~\cite{Chatrchyan:2012jua, CMS-PAS-BTV-15-001}. A ``medium'' working point with an efficiency of 70\% for genuine b jets and a rejection of 99\% of light-flavor jets is used, together with a ``loose'' working point that has an 85\% identification efficiency and rejects 90\% of light-flavored jets. The efficiency for identifying W, t, and b jets in simulation is corrected to match the results found in data.

An interesting feature of the direct production of a single vector-like T quark is the presence of an additional jet that is produced in the forward direction. Forward jets are reconstructed as AK4 jets using the same selections and corrections as defined above, but have $ 2.4 < \abs{\eta} < 5.0$ and $ \pt > 30\GeV$.

\section{Event selection}\label{sec:selection}
Events are required to have two oppositely charged leptons (either muons or electrons) forming a Z boson with an invariant mass between 70 and 110\GeV. A t quark from a T quark decay can be identified in three different ways: fully merged (a t jet is identified), partially merged (a W jet and a b jet are identified), or resolved (three AK4 jets are reconstructed). We therefore define ten event categories, depending on how the Z boson or the t quark candidates are reconstructed and on the number of forward jets present, as summarized in Table~\ref{tab:categories}.
\begin{table}[!h]
\centering
\topcaption{Summary of the ten categories of the analysis. For each category the leading lepton must have $ \pt > 120\GeV$, while at least one b jet has to be present.}\label{tab:categories}
\ifthenelse{\boolean{cms@external}}{\resizebox{\columnwidth}{!}}{}
{
\begin{tabular}{c c c c c c c}
\hline
Category & Z boson       & t quark          & $N$(forward jets) & $\Delta R (\ell,\ell)$ & $m_{\text{j1,j2}}$\\\hline
1        & two muons     & fully merged     & $\geq$0           & $<$1.4                 & \NA\\
2        & two electrons & fully merged     & $\geq$0           & $<$1.4                 & \NA\\
3        & two muons     & partially merged & 0                 & $<$0.6                 & \NA\\
4        & two muons     & partially merged & $\geq$1           & $<$0.6                 & \NA\\
5        & two electrons & partially merged & 0                 & $<$0.6                 & \NA\\
6        & two electrons & partially merged & $\geq$1           & $<$0.6                 & \NA\\
7        & two muons     & resolved         & 0                 & $<$0.6                 & $<$200\GeV\\
8        & two muons     & resolved         & $\geq$1           & $<$0.6                 & $<$200\GeV\\
9        & two electrons & resolved         & 0                 & $<$0.6                 & $<$200\GeV\\
10       & two electrons & resolved         & $\geq$1           & $<$0.6                 & $<$200\GeV\\\hline
\end{tabular}
}
\end{table}

The hierarchy places the most sensitive categories first. If an event falls into two or more categories it is assigned only to the first. For categories 1 and 2, the t quark candidate is given by the t jet; for categories 3--6 it is reconstructed by summing the momentum vectors of the W jet and the b jet; while for categories 7--10 the momenta of the three jets are summed. If more than one t quark candidate is found, the one with the largest \pt is selected for the subsequent restoration.

In addition to requiring a Z boson and a t quark in the event, at least one ``medium'' b jet has to be present (for the partially merged and the resolved categories, it is the one used to reconstruct the t quark), the two leptons from the Z boson decay have to be close to each other ($\Delta R < 0.6$--1.4, depending on the category), and the leading lepton (muon or electron) must have $ \pt > 120\GeV$. If more than one medium b jet is present, the one giving the largest t quark \pt is selected for subsequent reconstruction. Furthermore, in the resolved categories, the two jets with the lowest b tagging discriminant of the three jets forming the t quark candidate are required to have a dijet invariant mass $m_{\text{j1,j2}}$ below 200\GeV. All these requirements were optimized to increase the sensitivity of the analysis and are summarized in Table~\ref{tab:categories}.

The T quark candidate mass $m_{\text{tZ}}$ is obtained by summing the momenta of the Z candidate, given by the two muons or the two electrons, and the t quark candidate, reconstructed for the three scenarios as described above.

\section{Background estimate}\label{sec:background}
In this analysis, the signal is searched for as an excess in the mass spectrum of reconstructed T quark candidates, $m_{\text{tZ}}$, which is used as the discriminating variable. The background is largely dominated by Z/$\gamma$*+jets events (${>}80\%$), with smaller contributions from other sources (\ttbar+V, tZq, \ttbar, single t quark, and VV diboson production, where V represents a W or Z boson).

The background is estimated from data in order to reduce dependence on the simulation. This estimate, which incorporates all of the background processes described above, is obtained by measuring the $m_{tZ}$ distribution in a control region defined by applying the event selection described in Section 4, but instead of requiring the presence of a jet passing the medium b tagging requirements, a veto is applied on the presence of any jet passing the loose b tagging requirements. This veto effectively removes the signal while leaving a substantial fraction of the dominant Z+jets background.

The background expectation in the signal region is then estimated as:
\begin{equation}\label{eq:alphaRatioMethod}
N_{\text{bkg}}(\text{m}_{\text{tZ}}) = N_{\text{CR}}(\text{m}_{\text{tZ}}) \, \alpha(\text{m}_{\text{tZ}}),
\end{equation}
where $N_{\text{CR}}(\text{m}_{\text{tZ}})$ is the number of events found in the data in the control region as a function of $m_{\text{tZ}}$, and $\alpha(\text{m}_{\text{tZ}})$ is the ratio obtained from simulation of the number of background events in the signal region to that in the control region, at each value of $m_{\text{tZ}}$. A closure test is performed to validate the method in an independent signal-free region, defined by considering the resolved categories and inverting the cut on $m_{\text{j1,j2}}$. This region has been chosen because it has a negligible signal contamination and yet it preserves the background composition of the signal region. Good agreement is found between the predicted background and the observed data in this region, showing the robustness of the background estimation method. Furthermore a good agreement is also found between the predicted background using the described method and the predicted background from the simulated events.

\begin{figure*}[!h]
\centering
\includegraphics[width=0.47\textwidth]{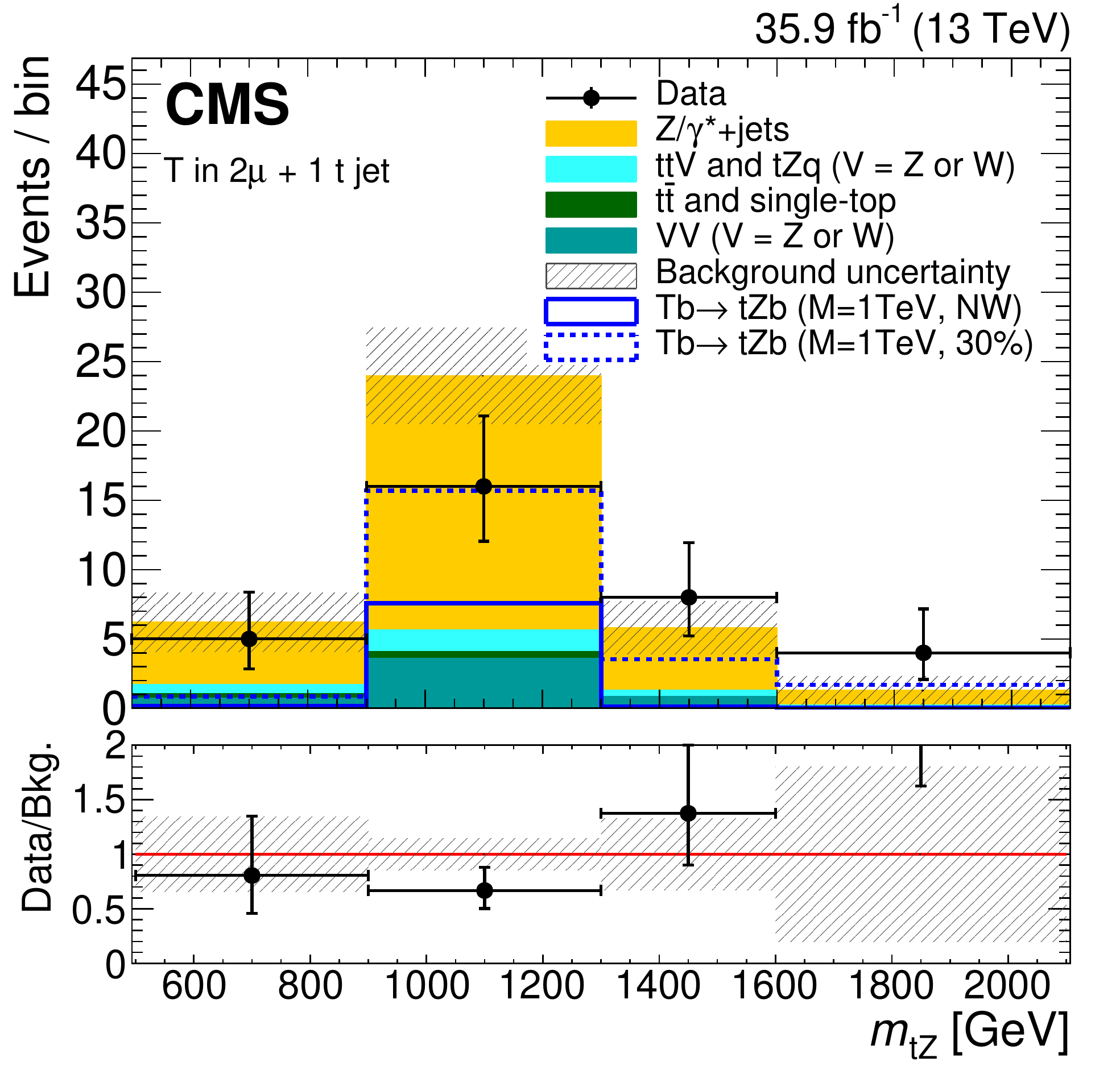}
\includegraphics[width=0.47\textwidth]{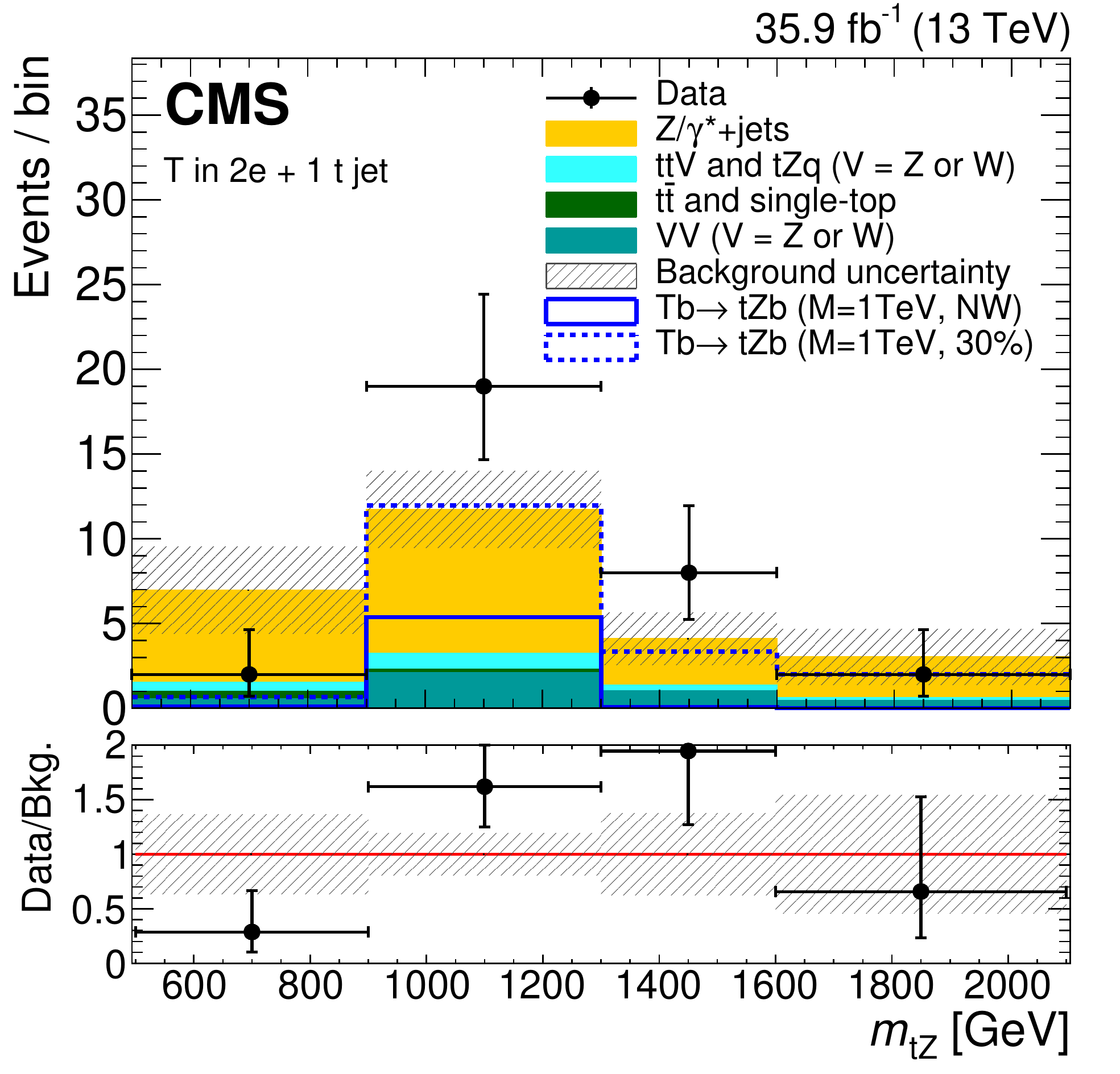}
\caption{Comparison between the data, the background estimate, and the expected signal for the 2 categories where the T quark is reconstructed in the fully merged topology, for events with the Z boson decaying into muons (left) and electrons (right). The background composition is taken from simulation. The uncertainties in the background estimate include both statistical and systematic components. The expected signal is shown for two benchmark values of the width, for a T quark produced in association with a b, T(b): narrow-width approximation (NW) and 30\% of the T quark mass. The lower panel in each plot shows the ratio of the data and the background estimation, with the shaded band representing the uncertainties in the background estimate. The vertical bars for the data points show the Poisson errors associated with each bin, while the horizontal bars indicate the bin width.}\label{fig:Background1}
\end{figure*}

\begin{figure*}[!h]
\centering
\includegraphics[width=0.47\textwidth]{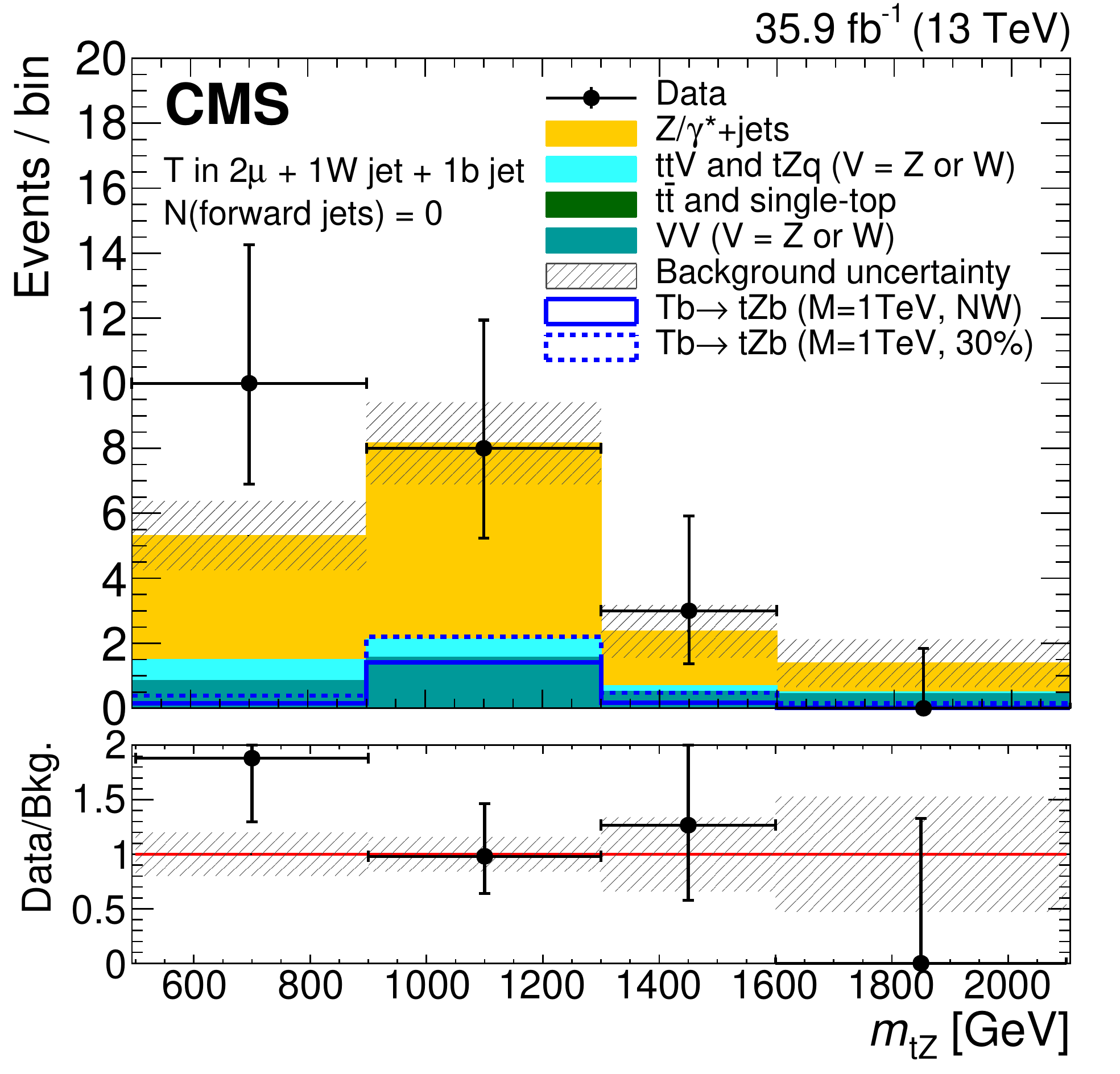}
\includegraphics[width=0.47\textwidth]{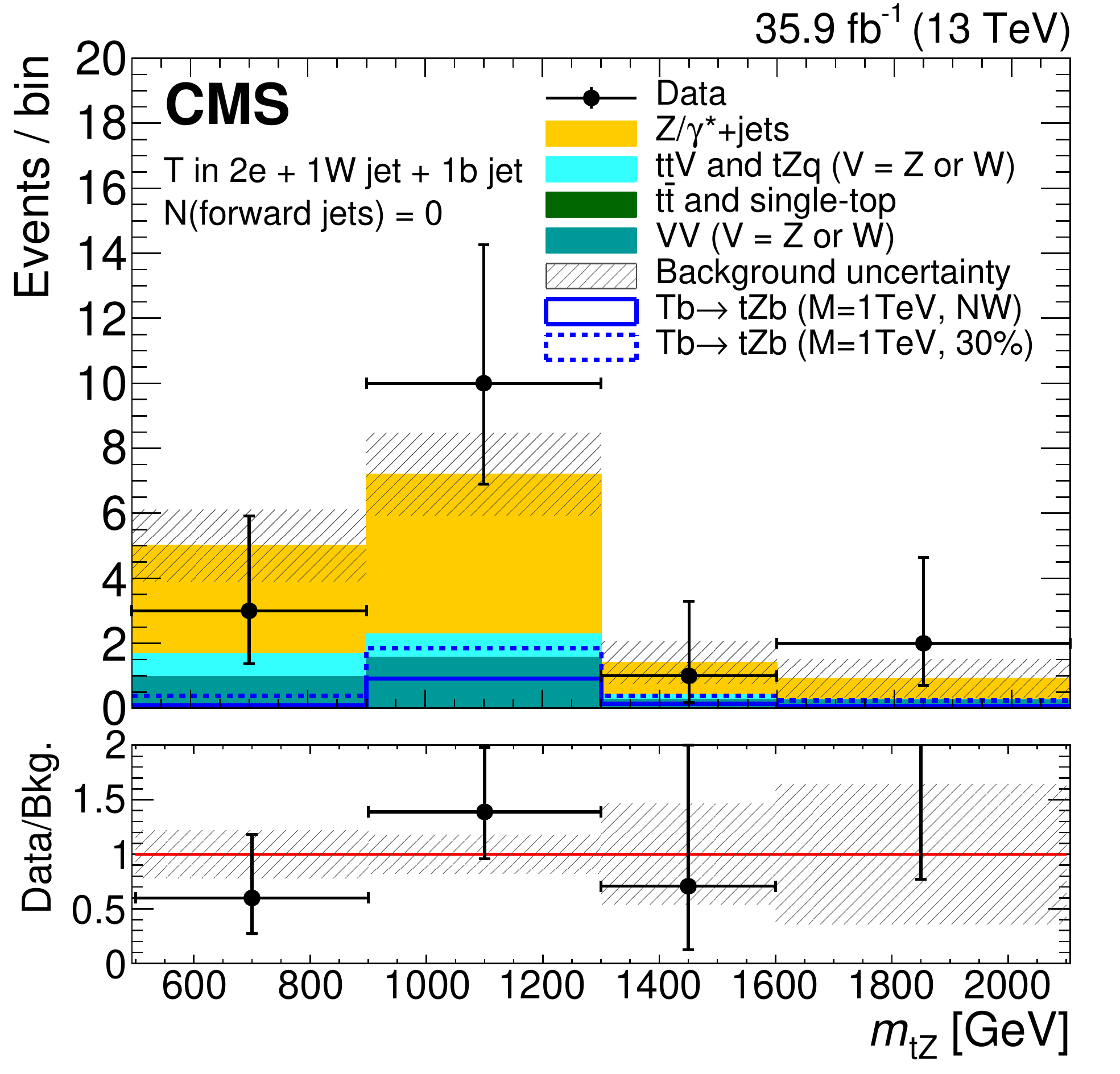}\\
\includegraphics[width=0.47\textwidth]{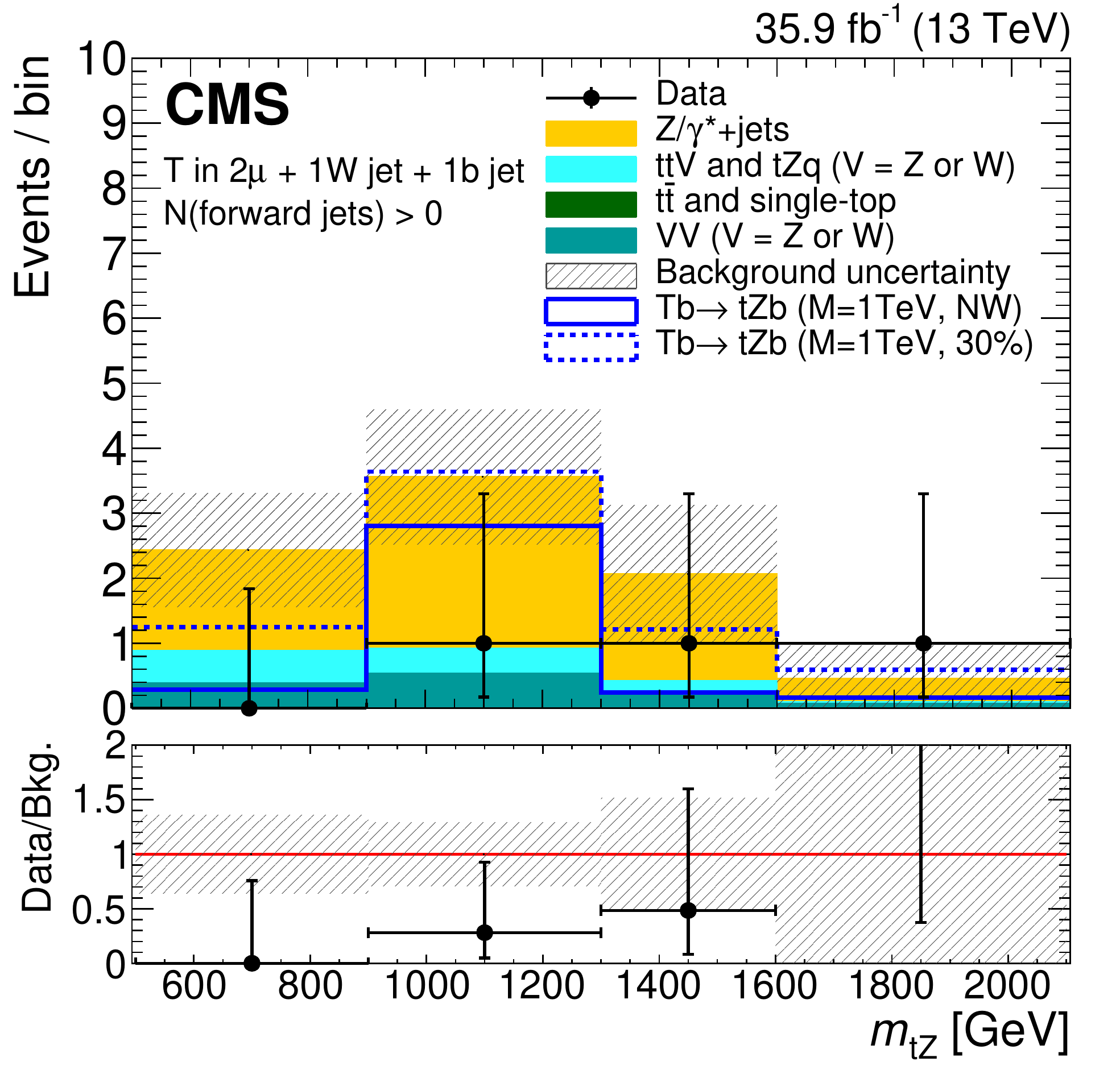}
\includegraphics[width=0.47\textwidth]{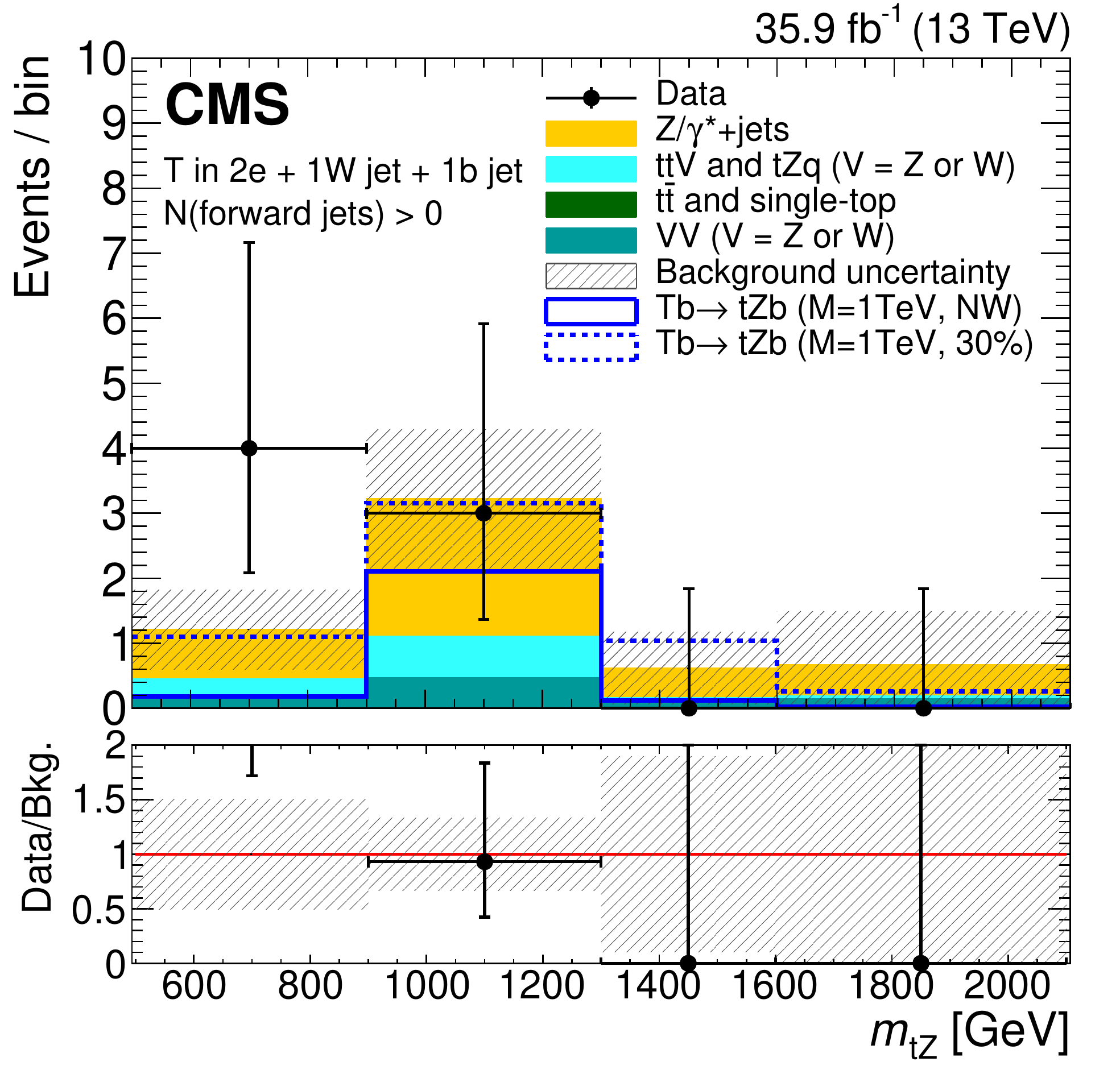}
\caption{Comparison between the data, the background estimate, and the expected signal for the 4 categories where the T quark is reconstructed in the partially merged topology, for events with the Z boson decaying into muons (left) and electrons (right), and zero (at least one) forward jets in the upper (lower) row. The background composition is taken from simulation. The uncertainties in the background estimate include both statistical and systematic components. The expected signal is shown for two benchmark values of the width, for a T quark produced in association with a b, T(b): narrow-width approximation (NW) and 30\% of the T quark mass. The lower panel in each plot shows the ratio of the data and the background estimation, with the shaded band representing the uncertainties in the background estimate. The vertical bars for the data points show the Poisson errors associated with each bin, while the horizontal bars indicate the bin width.}\label{fig:Background2}
\end{figure*}

\begin{figure*}[!h]
\centering
\includegraphics[width=0.47\textwidth]{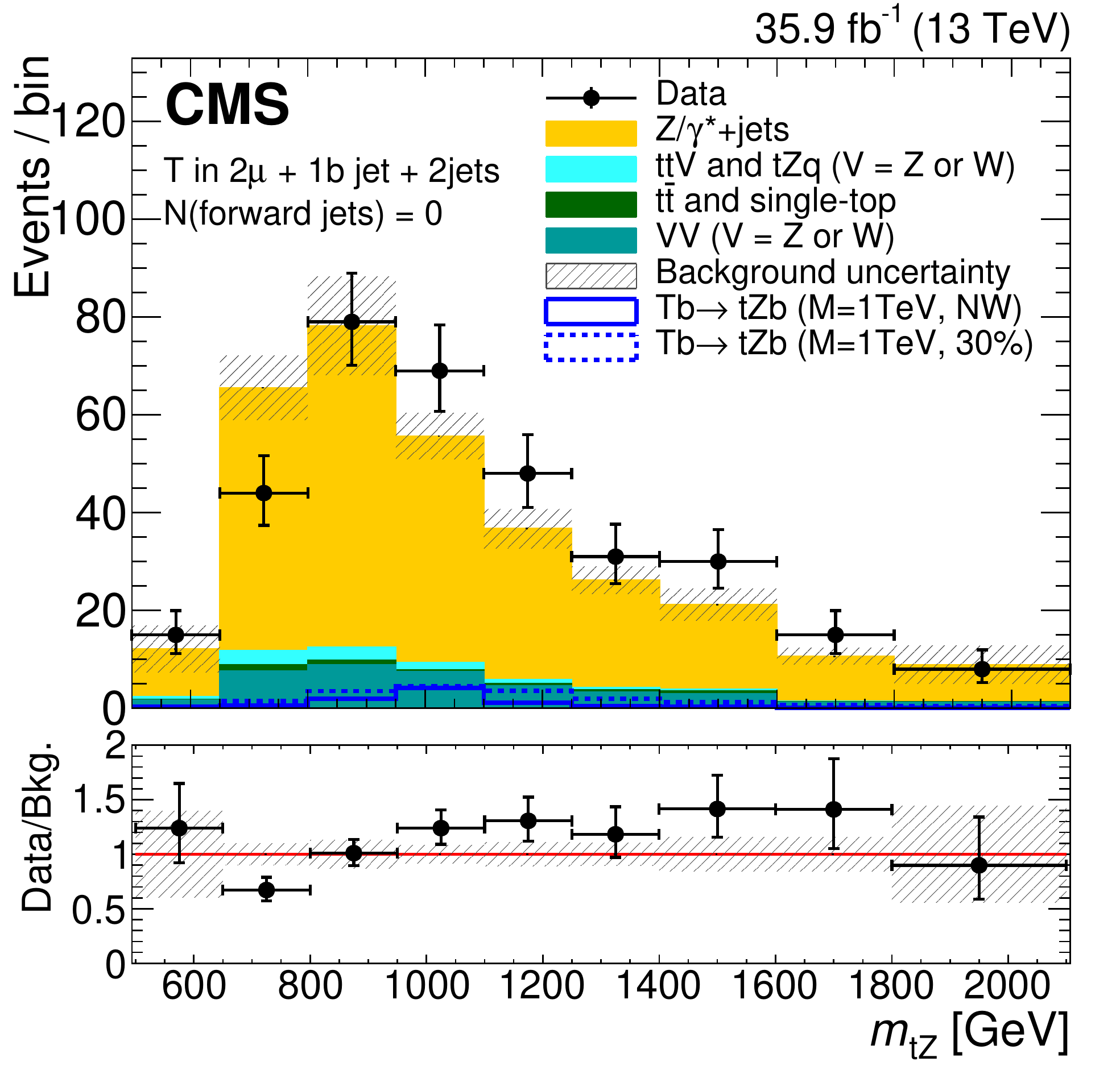}
\includegraphics[width=0.47\textwidth]{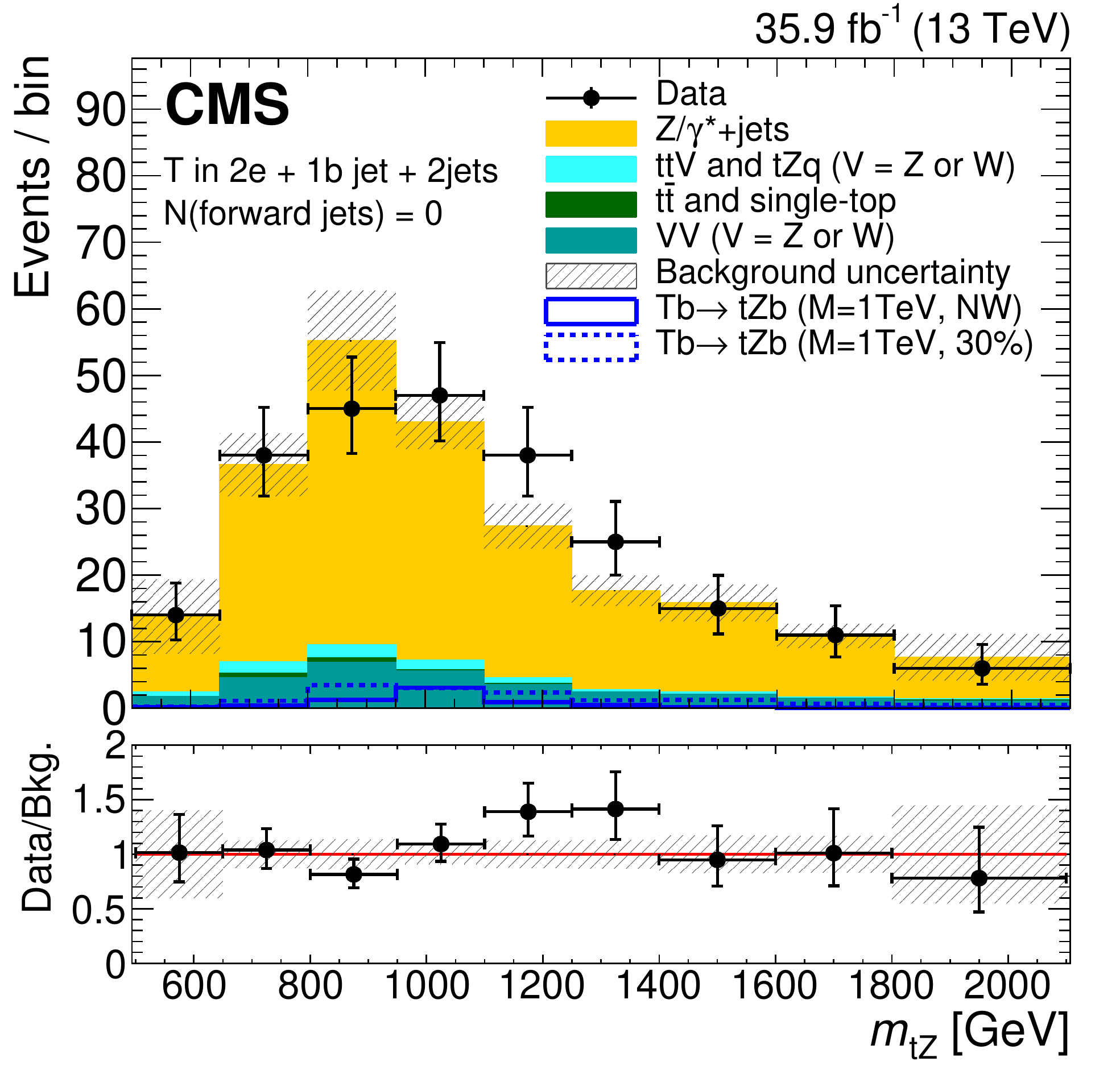}\\
\includegraphics[width=0.47\textwidth]{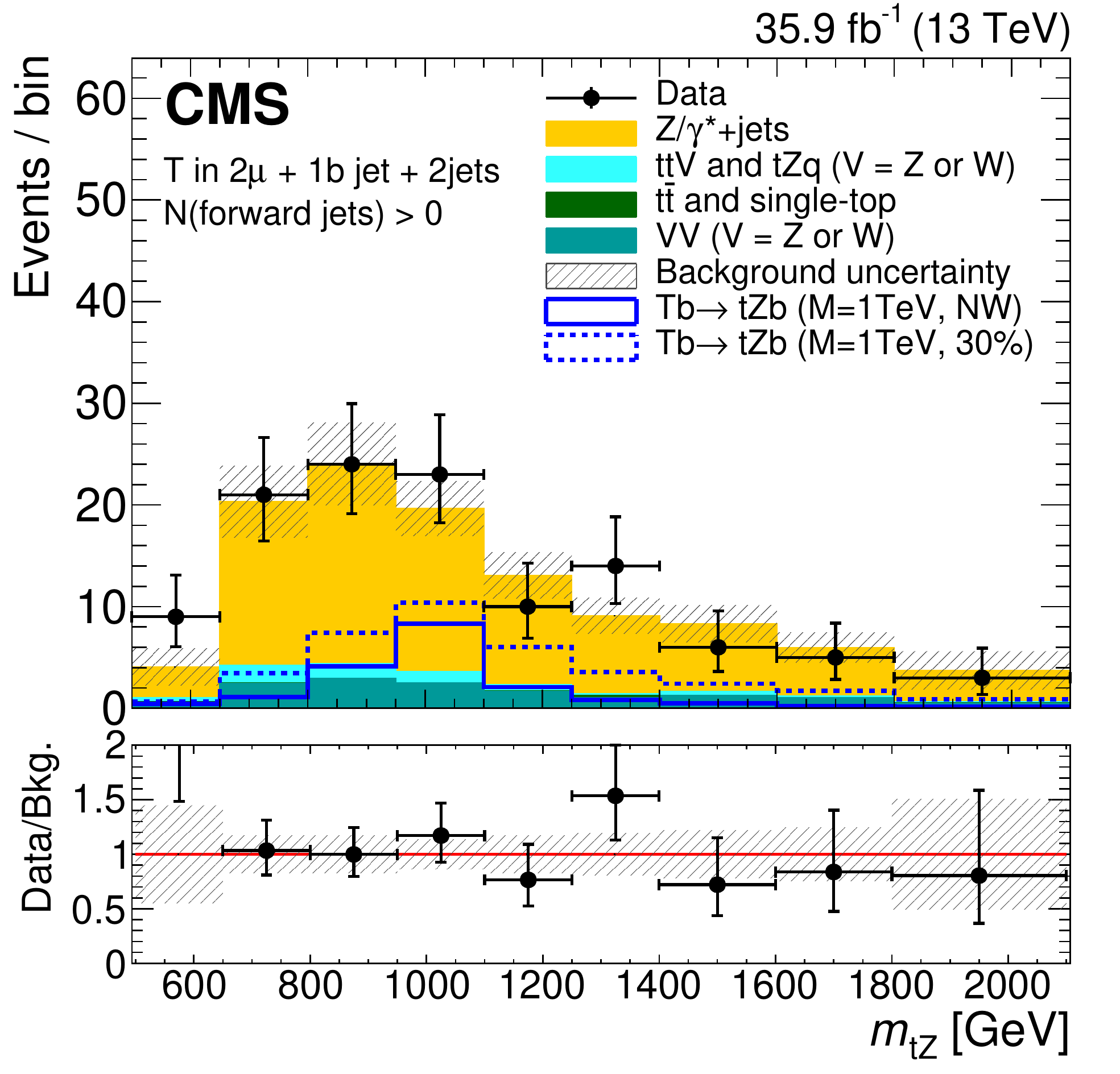}
\includegraphics[width=0.47\textwidth]{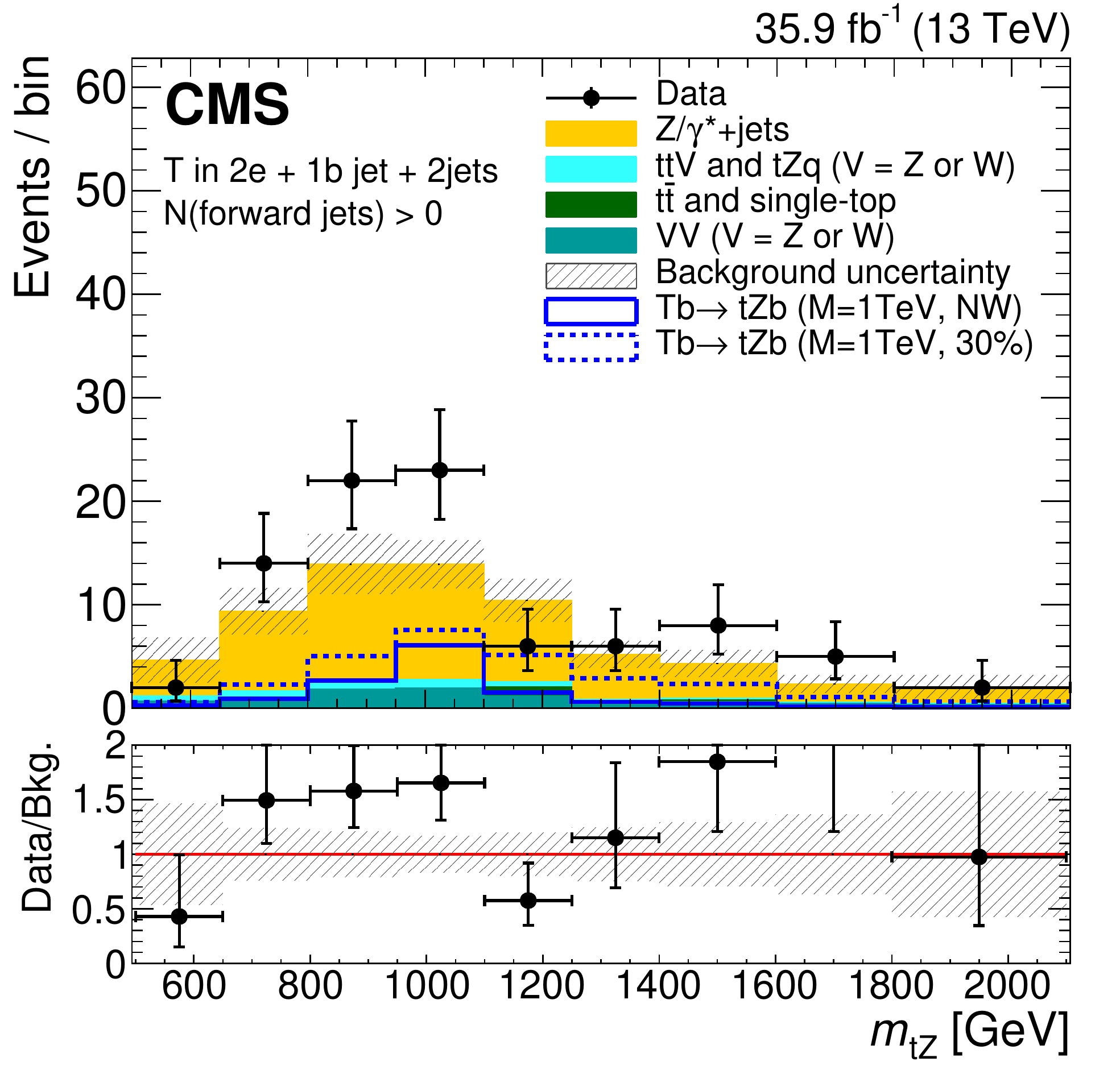}
\caption{Comparison between the data, the background estimate, and the expected signal for the 4 categories where the T quark is reconstructed in the resolved topology, for events with the Z boson decaying into muons (left) and electrons (right), and zero (at least one) forward jets in the upper (lower) row. The background composition is taken from simulation. The uncertainties in the background estimate include both statistical and systematic components. The expected signal is shown for two benchmark values of the width, for a T quark produced in association with a b, T(b): narrow-width approximation (NW) and 30\% of the T quark mass. The lower panel in each plot shows the ratio of the data and the background estimation, with the shaded band representing the uncertainties in the background estimate. The vertical bars for the data points show the Poisson errors associated with each bin, while the horizontal bars indicate the bin width.}\label{fig:Background3}
\end{figure*}

Comparisons between the background estimates and the observations in data in the $m_{\text{tZ}}$ distribution are shown in Figs.~\ref{fig:Background1},~\ref{fig:Background2}, and~\ref{fig:Background3}. The number of predicted background events and the number of observed events are reported in Tables~\ref{tab:Estimation1}, \ref{tab:Estimation2}, and \ref{tab:Estimation3}, together with the number of expected signal events for two example masses. The numbers of observed events are consistent with SM background predictions.

\clearpage

\begin{table*}[!h]
\centering
\topcaption{The number of estimated background events compared to the observed number of events for the two fully merged categories. The quoted uncertainties in the background estimates include both statistical and systematic components, as described in Section~\ref{sec:systematics}. Expected signal yields and their respective efficiencies in parentheses are given for two benchmark masses and two values of the width ``$\Gamma$'', for a T quark produced in association with a b, T(b), and a T quark produced in association with a t, T(t). The signal efficiencies are calculated for events with the Z boson decaying to electrons or muons. Background, data, and signal yields are shown for the range in $m_{\text{tZ}}$ between 500 and 2100\GeV.}\label{tab:Estimation1}
\begin{tabular}{l c c}
\hline
Channel                  & $ 2\mu {+} 1\,\PQt\text{-jet}$ & $ 2\Pe {+} 1\,\PQt\text{-jet}$ \\\hline
Estimated background     & $ 37.3 \pm 4.6 $             & $ 25.8 \pm 4.1 $          \\
Data events              & 33                         & 31                      \\ \hline
T(b), $m_{\mathrm{T}}$ = 0.8\TeV, $\Gamma$ $\simeq$ 0       &  \x\y1.2  (0.2\%) &  \x\y0.9 (0.1\%) \\
T(b), $m_{\mathrm{T}}$ = 0.8\TeV, $\Gamma$ = 0.3$m_{\mathrm{T}}$ &  22.9 (1\%)\x   & 17.1 (1\%)\x   \\
T(t), $\, m_{\mathrm{T}}$ = 0.8\TeV, $\Gamma$ $\simeq$ 0       &  1.3  (1\%)   &  1.0 (1\%)   \\
T(t), $\, m_{\mathrm{T}}$ = 0.8\TeV, $\Gamma$ = 0.3$m_{\mathrm{T}}$ &  6.3  (2\%)   &  5.4 (2\%)   \\\hline
T(b), $m_{\mathrm{T}}$ = 1.6\TeV, $\Gamma$ $\simeq$ 0       &  2.9  (6\%)   &  2.6 (6\%)   \\
T(b), $m_{\mathrm{T}}$ = 1.6\TeV, $\Gamma$ = 0.3$m_{\mathrm{T}}$ &  5.3  (5\%)   &  4.8 (5\%)   \\
T(t), $\, m_{\mathrm{T}}$ = 1.6\TeV, $\Gamma$ $\simeq$ 0       & 0.8  (6\%)   &  0.7 (6\%)   \\
T(t), $\, m_{\mathrm{T}}$ = 1.6\TeV, $\Gamma$ = 0.3$m_{\mathrm{T}}$ & 1.5  (5\%)   &  1.4 (5\%)   \\\hline
\end{tabular}
\end{table*}

\begin{table*}[!h]
\centering
\topcaption{The number of estimated background events compared to the observed number of events for the four partially merged categories. The quoted uncertainties in the background estimates include both statistical and systematic components, as described in Section~\ref{sec:systematics}. Expected signal yields and their respective efficiencies in parentheses are given for two benchmark masses and two values of the width ``$\Gamma$'', for a T quark produced in association with a b, T(b), and a T quark produced in association with a t, T(t). The signal efficiencies are calculated for events with the Z boson decaying to electrons or muons. Background, data, and signal yields are shown for the range in $m_{\text{tZ}}$ between 500 and 2100\GeV.}\label{tab:Estimation2}
\resizebox{\textwidth}{!}
{
\begin{tabular}{l c c c c}
\hline
\multirow{2}[0]{*}{Channel} & $ 2\mu {+} 1\,\PW\text{-jet} {+} 1\,\PQb\text{-jet}$ & $ 2\Pe {+} 1\,\PW\text{-jet} {+} 1\,\PQb\text{-jet} $ & $ 2\mu {+} 1\,\PW\text{-jet} {+} 1\,\PQb\text{-jet} $ & $ 2\Pe {+} 1\,\PW\text{-jet} {+} 1\,\PQb\text{-jet} $ \\
& \multicolumn{2}{c}{$N$(forward jets) = 0}       & \multicolumn{2}{c}{$N$(forward jets) $>$ 0}\\\hline
Estimated background        & $ 17.2 \pm 2.0 $         & $ 14.5 \pm 1.9 $        & $ 8.5 \pm 1.8 $       & $ 5.7 \pm 1.6  $ \\
Data events                 & 21                    & 16                     & 3                   & 7             \\\hline
T(b), $m_{\mathrm{T}}$ = 0.8\TeV, $\Gamma$ $\simeq$ 0       & 2.7  (0.5\%)          & 1.7  (0.3\%)           &  5.4  (0.9\%)       &  4.3  (0.7\%) \\
T(b), $m_{\mathrm{T}}$ = 0.8\TeV, $\Gamma$ = 0.3$m_{\mathrm{T}}$ & 8.2  (0.5\%)          & 5.0  (0.3\%)           &  12.2 (0.8\%)\x       &  9.5  (0.6\%) \\
T(t), $\, m_{\mathrm{T}}$ = 0.8\TeV, $\Gamma$ $\simeq$ 0       & 0.9  (0.8\%)          & 0.8  (0.7\%)           &  2.0  (2\%)\x\y         &  1.5  (1\%)\x\y   \\
T(t), $\, m_{\mathrm{T}}$ = 0.8\TeV, $\Gamma$ = 0.3$m_{\mathrm{T}}$ & 2.8  (0.9\%)          & 2.1  (0.6\%)           &  4.7  (1\%)\x\y         &  3.9  (1\%)\x\y   \\\hline
T(b), $m_{\mathrm{T}}$ = 1.6\TeV, $\Gamma$ $\simeq$ 0       & 0.2  (0.3\%)          & 0.2  (0.3\%)           &  0.4  (0.9\%)       &  0.3  (0.6\%) \\
T(b), $m_{\mathrm{T}}$ = 1.6\TeV, $\Gamma$ = 0.3$m_{\mathrm{T}}$ & 0.4  (0.4\%)          & 0.3  (0.3\%)           &  0.7  (0.7\%)       &  0.6  (0.6\%) \\
T(t), $\, m_{\mathrm{T}}$ = 1.6\TeV, $\Gamma$ $\simeq$ 0       & 0.1  (0.7\%)          & 0.1  (0.5\%)           &  0.2  (1\%)\x\y         &  0.2  (1\%)\x\y   \\
T(t), $\, m_{\mathrm{T}}$ = 1.6\TeV, $\Gamma$ = 0.3$m_{\mathrm{T}}$ & 0.2  (0.7\%)          & 0.2  (0.6\%)           &  0.4  (1\%)\x\y         &  0.4  (1\%)\x\y   \\\hline
\end{tabular}
}
\end{table*}

\begin{table*}[!h]
\centering
\topcaption{The number of estimated background events compared to the observed number of events for the four resolved categories. The quoted uncertainties in the background estimates include both statistical and systematic components, as described in Section~\ref{sec:systematics}. Expected signal yields and their respective efficiencies in parentheses are given for two benchmark masses and two values of the width ``$\Gamma$'', for a T quark produced in association with a b, T(b), and a T quark produced in association with a t, T(t). The signal efficiencies are calculated for events with the Z boson decaying to electrons or muons. Background, data, and signal yields are shown for the range in $m_{\text{tZ}}$ between 500 and 2100\GeV.}\label{tab:Estimation3}
\ifthenelse{\boolean{cms@external}}{}{\resizebox{\textwidth}{!}}
{
\begin{tabular}{l c c c c}
\hline
\multirow{2}[0]{*}{Channel} & $ 2\mu {+} 1\,\PQb\text{-jet} {+} 2\,\text{jets} $ & $2\Pe {+} 1\,\PQb\text{-jet} {+} 2 \,\text{jets} $ & $ 2\mu {+} 1\,\PQb\text{-jet} {+} 2 \,\text{jets} $ & $ 2\Pe {+} 1\,\PQb\text{-jet} {+} 2 \,\text{jets} $ \\
& \multicolumn{2}{c}{$N$(forward jets) = 0}       & \multicolumn{2}{c}{$N$(forward jets) $>$ 0}\\\hline
Estimated background        & $ 315 \pm 16 $     & $ 228 \pm 13 $     & $ 108.3 \pm 7.5 $      & $ 66.2 \pm 5.7 $  \\
Data events                 & 339                  & 239                  & 115                  & 88              \\\hline
T(b), $m_{\mathrm{T}}$ = 0.8\TeV, $\Gamma$ $\simeq$ 0       & 13.7 (2\%)\x           & 10.0 (2\%)\x           & 25.7 (4\%)\x           & 18.5 (3\%)\x      \\
T(b), $m_{\mathrm{T}}$ = 0.8\TeV, $\Gamma$ = 0.3$m_{\mathrm{T}}$ & 35.9 (2\%)\x           & 29.7 (2\%)\x           & 66.5 (4\%)\x           & 52.7 (3\%)\x      \\
T(t), $\,m_{\mathrm{T}}$ = 0.8\TeV, $\Gamma$ $\simeq$ 0       & 2.5  (2\%)           & 2.0  (2\%)           & 5.0  (5\%)           & 4.0  (4\%)      \\
T(t), $\,m_{\mathrm{T}}$ = 0.8\TeV, $\Gamma$ = 0.3$m_{\mathrm{T}}$ & 8.9  (3\%)           & 6.7  (2\%)           & 15.8 (5\%)\x           & 12.0 (4\%)\x      \\\hline
T(b), $m_{\mathrm{T}}$ = 1.6\TeV, $\Gamma$ $\simeq$ 0       & 1.0  (2\%)           & 0.9  (2\%)           & 2.5  (5\%)           & 2.0  (4\%)      \\
T(b), $m_{\mathrm{T}}$ = 1.6\TeV, $\Gamma$ = 0.3$m_{\mathrm{T}}$ & 2.2  (2\%)           & 1.9  (2\%)           & 4.7  (5\%)           & 3.9  (4\%)      \\
T(t), $\,m_{\mathrm{T}}$ = 1.6\TeV, $\Gamma$ $\simeq$ 0       & 0.3  (3\%)           & 0.3  (2\%)           & 0.8  (6\%)           & 0.7  (5\%)      \\
T(t), $\,m_{\mathrm{T}}$ = 1.6\TeV, $\Gamma$ = 0.3$m_{\mathrm{T}}$ & 0.8  (3\%)           & 0.7  (2\%)           & 1.7  (6\%)           & 1.5  (5\%)      \\\hline
\end{tabular}
}
\end{table*}
\clearpage

\section{Systematic uncertainties}\label{sec:systematics}

Systematic effects have been evaluated by propagating the uncertainties in the input quantities. Unless explicitly stated, the impact of these uncertainties are evaluated both in the normalization and in the shape of the distribution of $m_{\text{tZ}}$.

Five main sources of uncertainty contribute to the estimated background. The dominant ones are the statistical uncertainties in the control regions used to estimate the background, both in data, giving an uncertainty of 10--46\% depending on the category, and in the simulation, with an uncertainty of 3--34\%. The small differences between the observation and the prediction for the closure test described previously are taken as systematic uncertainties (6\%). An uncertainty due to possible mismodelling of the Z+light quark and Z+b quark fractions in the simulation is evaluated. This systematic uncertainty is evaluated by observing the effect of changing the Z+b fraction by 10\%~\cite{Chatrchyan:2014dha}, yielding a contribution to the uncertainty in the background estimation of between 2 and 4\%. Finally, the uncertainty from the b tagging for the b, c, and light-flavor jets is evaluated by changing the b tagging corrections by their uncertainties~\cite{Chatrchyan:2012jua, CMS-PAS-BTV-15-001}, yielding a change in the normalization of 2\% for the b tagging efficiency and 2\% for the misidentification probability.
Other systematic uncertainties related to the simulation modelling have been studied and found negligible, because of the data-driven method used to estimate the background.

The systematic uncertainty in the signal is estimated from the corrections applied to the simulation to match distributions in data. The corrections for lepton identification and lepton trigger efficiency are obtained from dedicated analyses, using the ``tag-and-probe'' method~\cite{Chatrchyan:2012xi, Khachatryan:2015hwa}. Changing these corrections by their uncertainties provides an estimate of the uncertainties in the signal yield of 3\% for muons and electrons for a mass hypothesis of 1.0\TeV, and 1\% for the trigger. The jet four-momenta are varied by the JES and JER uncertainties, which provide respective changes in the signal yield of 1\% (JES) and 0.5\% (JER), while for forward jets a change of 8\% is observed. For W and t jet tagging, the same procedure of varying the corrections is applied, yielding an uncertainty of 4 and 8\%, respectively. The uncertainty in the b tagging efficiency is evaluated, as for the background; the change in yield of the signal is found to be 2.5\%. The uncertainties from the choice of PDF are evaluated using the {\sc nnpdf}~3.0 PDF eigenvectors~\cite{Butterworth:2015oua}, considering only the change in the shape of the $m_{\text{tZ}}$ distribution. The uncertainty in the simulation of pileup is obtained by changing the inelastic cross section, which controls the average pileup multiplicity, by 5\%~\cite{Aaboud:2016mmw}, resulting in a signal yield uncertainty of 1\%. Additional sources of systematic uncertainty are the integrated luminosity (2.5\%, normalization only)~\cite{CMS-PAS-LUM-17-001} and the factorization and renormalization scales used in simulation (shape only).

\section{Results}\label{sec:results}

No significant deviations from the expected background are observed in any of the search channels. We set upper limits on the product of the cross section and branching fraction of a T quark decaying to tZ. The exclusion limits at a confidence level (CL) of 95\% are obtained using the asymptotic CL$_\mathrm{s}$ criterion~\cite{Read:2002hq, Junk:1999kv, Cowan:2010js, ATLAS:2011tau}, with templates for background and signal given by the binned distributions in Figs.~\ref{fig:Background1},~\ref{fig:Background2}, and~\ref{fig:Background3}. Systematic uncertainties are treated as nuisance parameters, assuming a log-normal distribution for normalization parameters and a Gaussian distribution for systematic uncertainties that affect the $m_{\text{tZ}}$ shape.

\begin{figure*}[!h]
\centering
\includegraphics[width=0.47\textwidth]{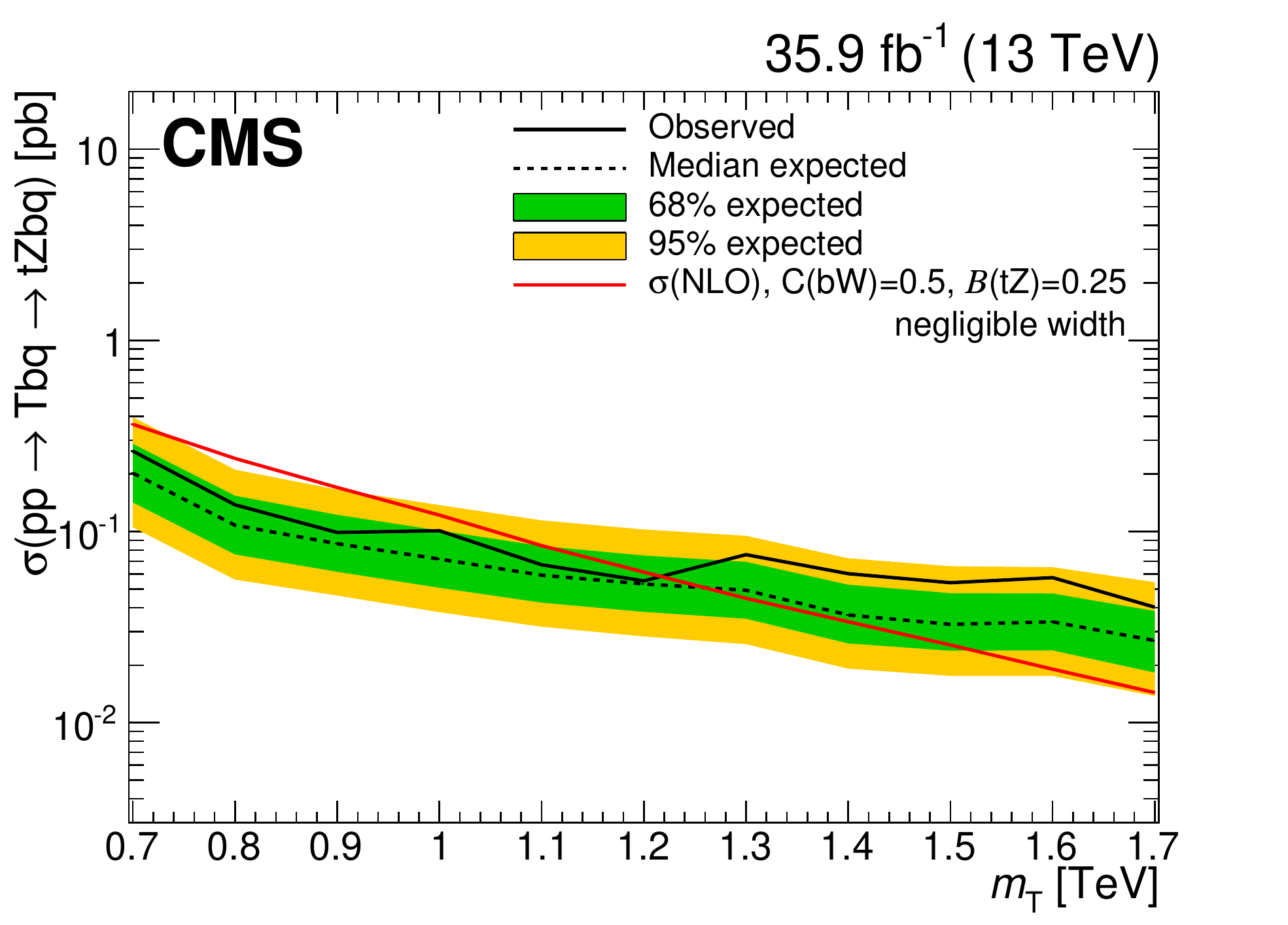}
\includegraphics[width=0.47\textwidth]{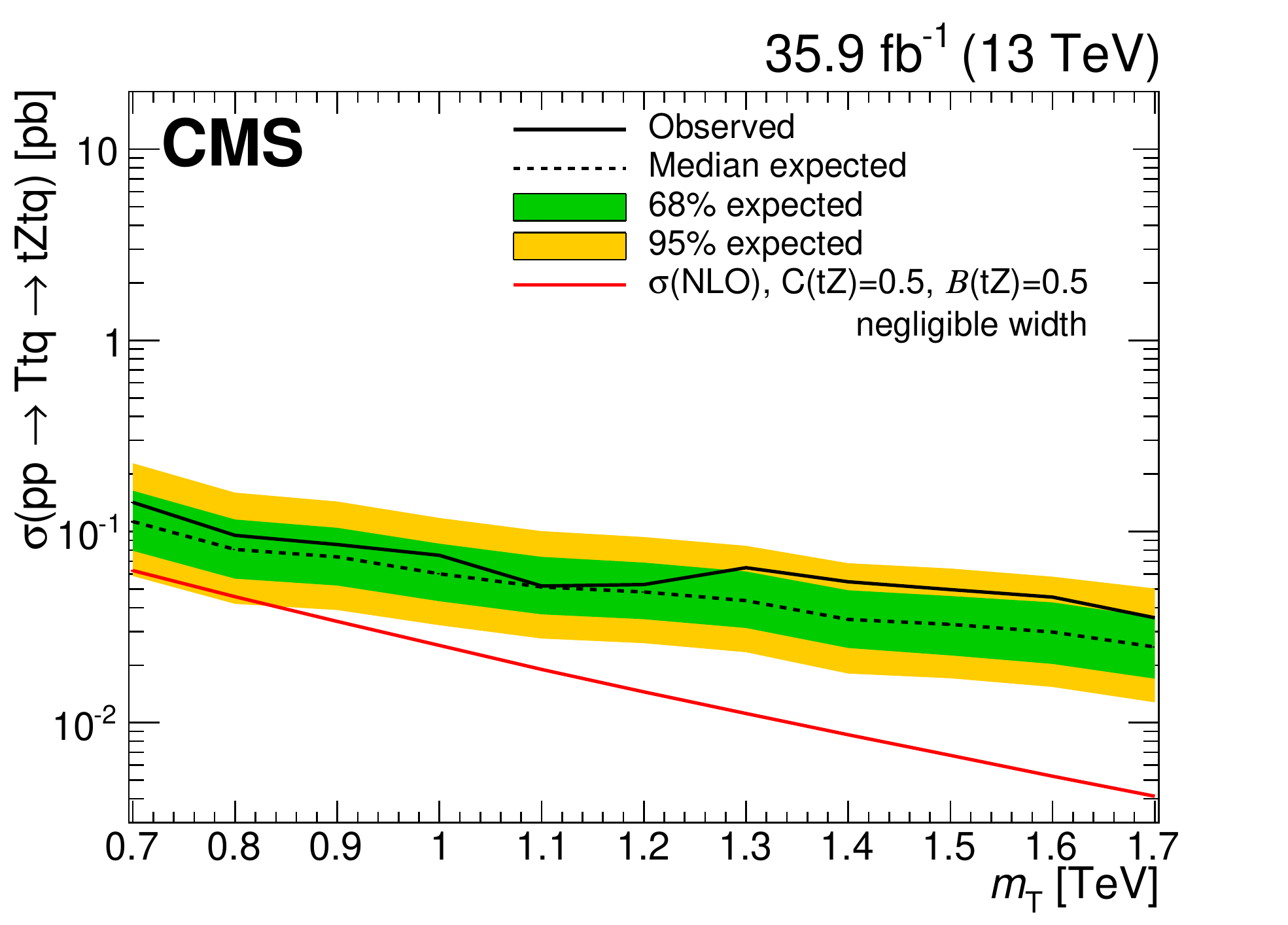}
\caption{Observed and expected limits at 95\% CL on the product of the single production cross section and branching fraction for the singlet LH T quark produced in association with a b quark (left) and for the doublet RH T quark produced in association with a t quark (right), where the T quark has a narrow width and decays to tZ. The inner green and outer yellow bands represent the 1 and 2 standard deviation uncertainties in the expected limit. The red lines indicate theoretical cross sections, as calculated at next-to-leading order in Ref. [4]. The branching fraction $\mathcal{B}$(T$\to$ tZ) is 0.25 (0.5) for the left (right) plot.}\label{fig:limitTprime1}
\end{figure*}

In Fig.~\ref{fig:limitTprime1}, the observed and expected limits from the ten categories of the T quark search are shown combined together, for the singlet LH T(b) (left) and doublet RH T(t) (right) production modes. The ten categories have different sensitivities to different values of $m_{\mathrm{T}}$, and the final result benefits from this behavior: the resolved categories drive the limit at low $m_{\mathrm{T}}$, the fully merged categories, at higher values, while at intermediate values the limit takes advantage of all the three topologies. Limits on $\sigma(\Pp\Pp \to \mathrm{Tbq} \to \mathrm{tZbq})$ for the singlet LH T(b) exclude values greater than 0.26--0.04\unit{pb} at 95\%~CL, for masses in the range 0.7--1.7\TeV. For an RH T(t) signal, the region above 0.14--0.04\unit{pb} is excluded for the same mass range. Upper limits are compared with theoretical cross sections calculated at NLO in Ref.~\cite{Matsedonskyi:2014mna}. For this model, a singlet LH T quark with C(bW) = 0.5 is excluded at 95\%~CL for masses in the range 0.7--1.2\TeV.

\begin{figure*}[!h]
\centering
\includegraphics[width=0.47\textwidth]{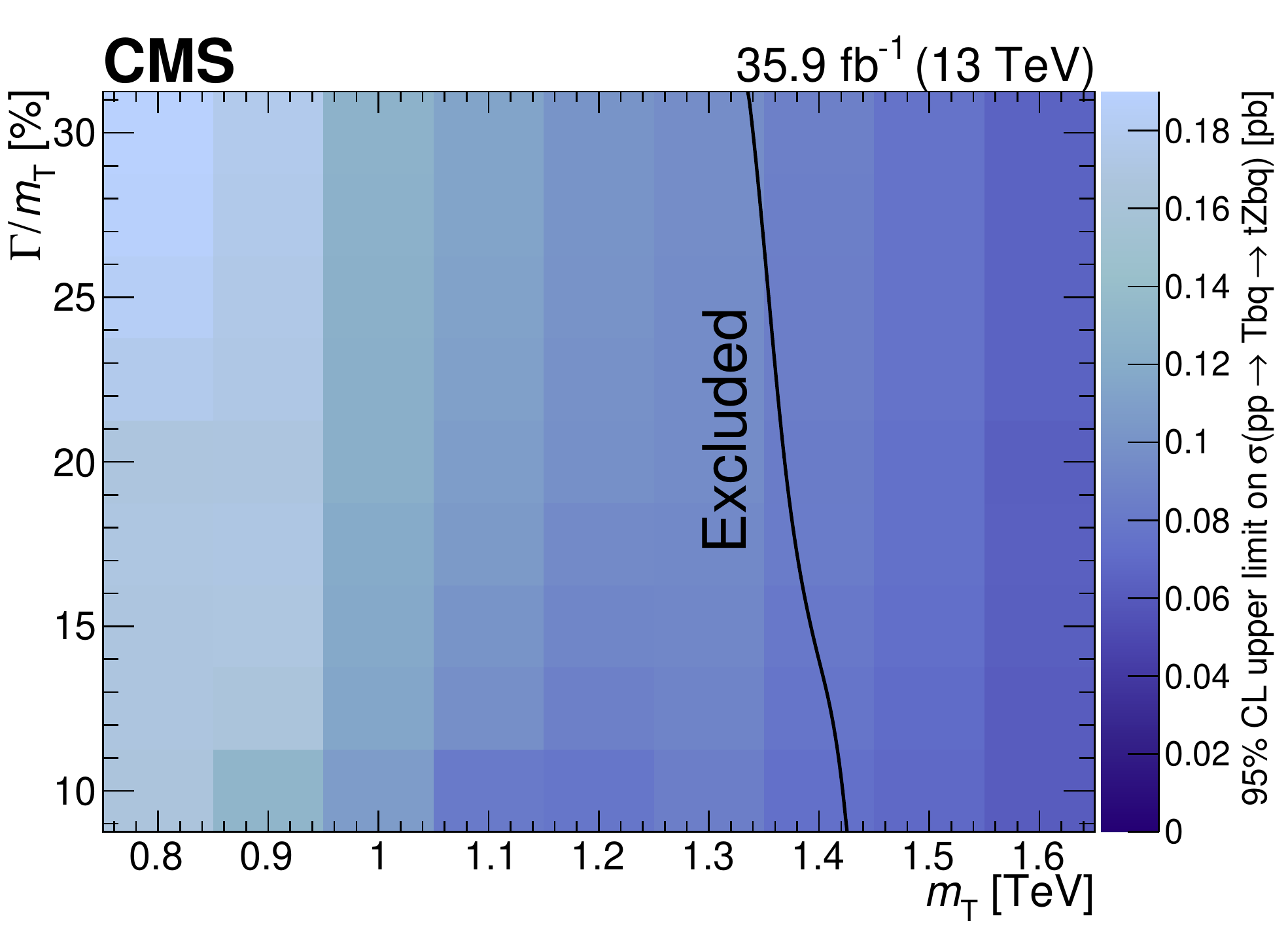}
\includegraphics[width=0.47\textwidth]{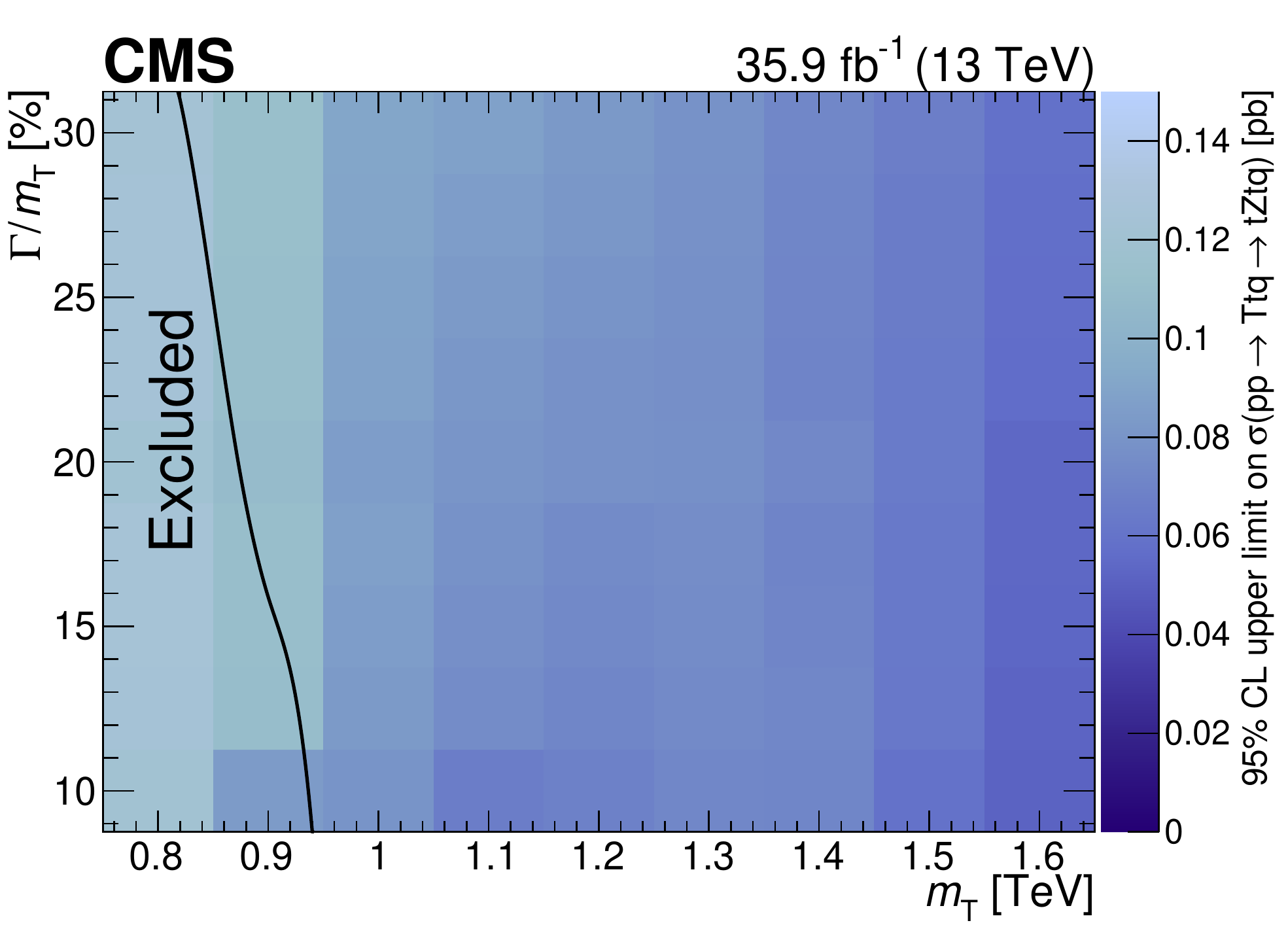}\\
\includegraphics[width=0.47\textwidth]{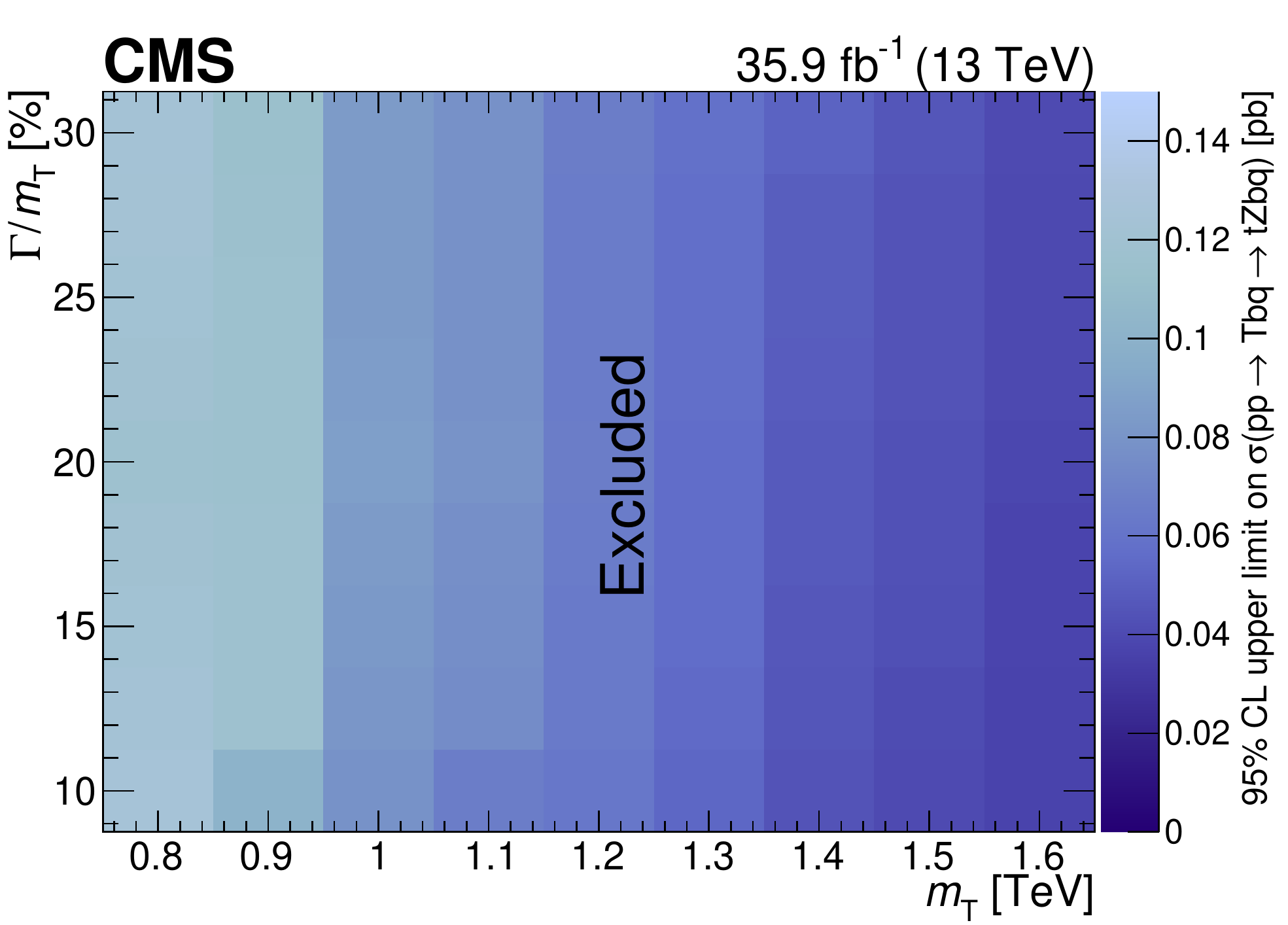}
\includegraphics[width=0.47\textwidth]{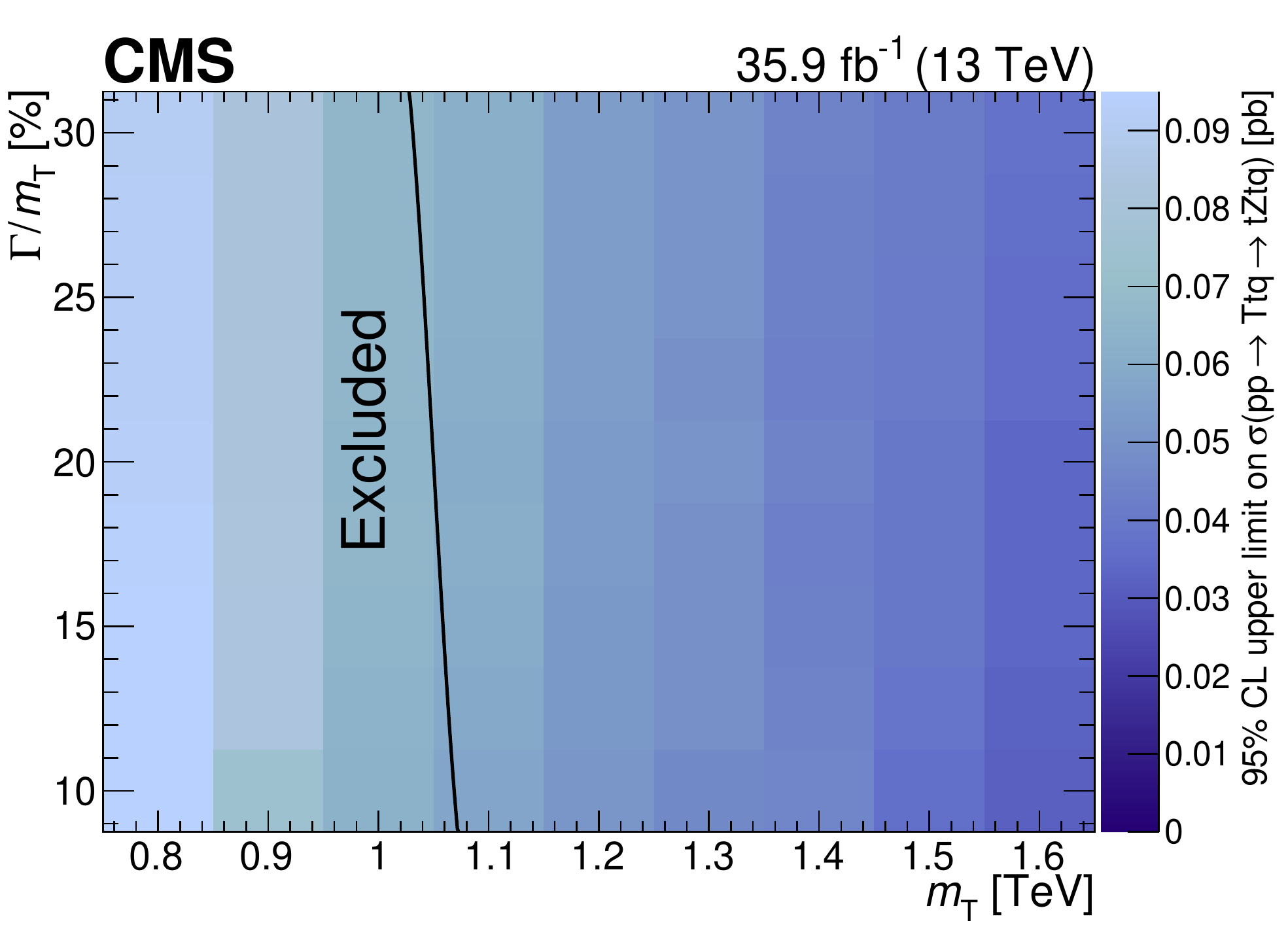}
\caption{Observed (upper) and expected (lower) limits at 95\% CL on the product of the single production cross section and branching fraction for the singlet LH T quark produced in association with a b quark (left) and for the doublet RH T quark produced in association with a t quark (right), where the T quark has a width from 10\% to 30\% of its mass and decays to tZ. The solid black lines indicate theoretical cross sections, as calculated at leading order using a modified version of the model constructed by the authors of Refs.~\cite{Buchkremer:2013bha,Fuks:2016ftf,Oliveira:2014kla} and reported in Table ~\ref{tab:CrossSec2}. In each plot, the excluded region lies to the left of the line, except in the lower-left plot where the entire region shown is excluded.}\label{fig:limitTprime2}
\end{figure*}

In Fig.~\ref{fig:limitTprime2}, the observed and expected upper limits at 95\% CL are shown as a function of the T quark width and T quark mass in the ranges from 10 to 30\% and 0.8 to 1.6\TeV, respectively. A sensitivity similar to that obtained assuming a narrow-width T quark is observed. In this case the experimental results are compared with the theoretical cross sections calculated at LO using a modified version of the model constructed by the authors of~\cite{Buchkremer:2013bha,Fuks:2016ftf,Oliveira:2014kla} and reported in Table~\ref{tab:CrossSec2}. For this model, the data exclude a singlet LH T quark produced in association with a b quark, for masses below values in the range 1.34 and 1.42\TeV depending on the width. A doublet RH T quark produced in association with a t quark is excluded for masses below values in the range 0.82 and 0.94\TeV.

In addition to being singly produced directly, as diagrammed in Fig.~\ref{fig:feynman} (\cmsLeft), the T quark may also appear singly in events where a single \PZpr is produced that decays $\PZpr\to\mathrm{T}\PQt$, as illustrated in Fig.~\ref{fig:feynman} (\cmsRight). Observed and expected limits for the production of a T quark via the decay of a \PZpr boson, $\PZpr\to\mathrm{T}\PQt$ and $\mathrm{T}\to\PQt\Z$, are shown in Table~\ref{tab:Zprime}. We assume negligible widths for both the \PZpr boson and the T quark. The product of cross section and branching fractions is excluded above 0.13--0.06\unit{pb}, for a \PZpr boson mass in the range from 1.5 to 2.5\TeV and for a T quark mass from 0.7 to 1.5\TeV.

\begin{table*}[!h]
\centering
\topcaption{Observed and expected 95\%~CL upper limit on $\sigma(\Pp\Pp\to\PZpr)\,\mathcal{B}(\PZpr\to\mathrm{T}\PQt)\,\mathcal{B}(\mathrm{T}\to\PQt\Z)$. The $\pm$1 and $\pm$2 standard deviation (s.d.) expected limits are also given. The limits are given in~pb.}\label{tab:Zprime}
\begin{tabular}{c c c c c c }
\hline
$m_{\text{Z}^{\prime}}$ [\TeVns{}] & $m_{\mathrm{T}}$ [\TeVns{}] & Observed & Expected & Expected $-$1(2) s.d. & Expected $+$1(2) s.d. \\
\hline
1.5 & 0.7 & 0.13 & 0.10 & 0.07 (0.05) & 0.14 (0.19) \\
1.5 & 0.9 & 0.11 & 0.08 & 0.06 (0.05) & 0.12 (0.16) \\
1.5 & 1.2 & 0.09 & 0.05 & 0.04 (0.03) & 0.07 (0.10) \\
2.0 & 0.9 & 0.08 & 0.06 & 0.04 (0.03) & 0.08 (0.11) \\
2.0 & 1.2 & 0.08 & 0.05 & 0.04 (0.03) & 0.07 (0.09) \\
2.0 & 1.5 & 0.06 & 0.04 & 0.03 (0.02) & 0.05 (0.07) \\
2.5 & 1.2 & 0.06 & 0.05 & 0.03 (0.02) & 0.06 (0.09) \\
2.5 & 1.5 & 0.06 & 0.04 & 0.03 (0.02) & 0.05 (0.07) \\
\hline
\end{tabular}
\end{table*}
\clearpage

\section{Summary}\label{sec:summary}
This paper has presented results of a search for the single production of a T quark with a charge of $+2/3$, decaying to a Z boson and a t quark. No deviations were observed relative to the expected standard model background. Upper limits on the product of the cross section and branching fraction range between 0.26 and 0.04\unit{pb} at 95\% confidence level for a left-handed T quark produced in association with a b quark, T(b), and between 0.14 and 0.04\unit{pb} for a right-handed T quark produced in association with a t quark, T(t), for the range of masses between 0.7 and 1.7\TeV. This result was obtained under the hypothesis of a narrow-width T quark, providing an interpretation of results through the simplified approach of Ref.~\cite{Matsedonskyi:2014mna}. In this case, left-handed T quarks produced in association with a b quark and with a coupling C(bW) of 0.5 were excluded for masses in the range 0.7--1.2\TeV. A large gain in the search sensitivity was found relative to previous results~\cite{Sirunyan:2017ezy} because of improvements introduced in the analysis as well as the increase in the integrated luminosity. The effect of a nonnegligible width was also studied; values of the width between 10 and 30\% of the T quark mass were considered, and similar sensitivities were observed. The results were interpreted using a modified version of the model constructed by the authors of Refs.~\cite{Buchkremer:2013bha,Fuks:2016ftf,Oliveira:2014kla}, and a left-handed T(b) signal was excluded for masses below values in the range 1.34--1.42\TeV, depending on the width, while a right-handed T(t) signal was excluded for masses below values in the range 0.82--0.94\TeV. Finally, the production of a \PZpr boson that decays to Tt was excluded for values of the product of cross section and branching fractions between 0.13--0.06\unit{pb}, for \PZpr boson and T quark masses in the respective ranges of 1.5 to 2.5\TeV and 0.7 to 1.5\TeV. The results presented in this paper are the most-stringent limits to date on the single production of heavy vector-like T quarks, the first to set limits for a variety of resonance widths, and the most-stringent limits for the production of a \PZpr boson decaying to Tt.

\begin{acknowledgments}

We congratulate our colleagues in the CERN accelerator departments for the excellent performance of the LHC and thank the technical and administrative staffs at CERN and at other CMS institutes for their contributions to the success of the CMS effort. In addition, we gratefully acknowledge the computing centers and personnel of the Worldwide LHC Computing Grid for delivering so effectively the computing infrastructure essential to our analyses. Finally, we acknowledge the enduring support for the construction and operation of the LHC and the CMS detector provided by the following funding agencies: BMWFW and FWF (Austria); FNRS and FWO (Belgium); CNPq, CAPES, FAPERJ, and FAPESP (Brazil); MES (Bulgaria); CERN; CAS, MoST, and NSFC (China); COLCIENCIAS (Colombia); MSES and CSF (Croatia); RPF (Cyprus); SENESCYT (Ecuador); MoER, ERC IUT, and ERDF (Estonia); Academy of Finland, MEC, and HIP (Finland); CEA and CNRS/IN2P3 (France); BMBF, DFG, and HGF (Germany); GSRT (Greece); OTKA and NIH (Hungary); DAE and DST (India); IPM (Iran); SFI (Ireland); INFN (Italy); MSIP and NRF (Republic of Korea); LAS (Lithuania); MOE and UM (Malaysia); BUAP, CINVESTAV, CONACYT, LNS, SEP, and UASLP-FAI (Mexico); MBIE (New Zealand); PAEC (Pakistan); MSHE and NSC (Poland); FCT (Portugal); JINR (Dubna); MON, RosAtom, RAS, RFBR and RAEP (Russia); MESTD (Serbia); SEIDI, CPAN, PCTI and FEDER (Spain); Swiss Funding Agencies (Switzerland); MST (Taipei); ThEPCenter, IPST, STAR, and NSTDA (Thailand); TUBITAK and TAEK (Turkey); NASU and SFFR (Ukraine); STFC (United Kingdom); DOE and NSF (USA).

\hyphenation{Rachada-pisek} Individuals have received support from the Marie-Curie program and the European Research Council and Horizon 2020 Grant, contract No. 675440 (European Union); the Leventis Foundation; the A. P. Sloan Foundation; the Alexander von Humboldt Foundation; the Belgian Federal Science Policy Office; the Fonds pour la Formation \`a la Recherche dans l'Industrie et dans l'Agriculture (FRIA-Belgium); the Agentschap voor Innovatie door Wetenschap en Technologie (IWT-Belgium); the Ministry of Education, Youth and Sports (MEYS) of the Czech Republic; the Council of Science and Industrial Research, India; the HOMING PLUS program of the Foundation for Polish Science, cofinanced from European Union, Regional Development Fund, the Mobility Plus program of the Ministry of Science and Higher Education, the National Science Center (Poland), contracts Harmonia 2014/14/M/ST2/00428, Opus 2014/13/B/ST2/02543, 2014/15/B/ST2/03998, and 2015/19/B/ST2/02861, Sonata-bis 2012/07/E/ST2/01406; the National Priorities Research Program by Qatar National Research Fund; the Programa Clar\'in-COFUND del Principado de Asturias; the Thalis and Aristeia programs cofinanced by EU-ESF and the Greek NSRF; the Rachadapisek Sompot Fund for Postdoctoral Fellowship, Chulalongkorn University and the Chulalongkorn Academic into Its 2nd Century Project Advancement Project (Thailand); and the Welch Foundation, contract C-1845.

\end{acknowledgments}

\bibliography{auto_generated}

\cleardoublepage \appendix\section{The CMS Collaboration \label{app:collab}}\begin{sloppypar}\hyphenpenalty=5000\widowpenalty=500\clubpenalty=5000\vskip\cmsinstskip
\textbf{Yerevan Physics Institute,  Yerevan,  Armenia}\\*[0pt]
A.M.~Sirunyan,  A.~Tumasyan
\vskip\cmsinstskip
\textbf{Institut f\"{u}r Hochenergiephysik,  Wien,  Austria}\\*[0pt]
W.~Adam,  F.~Ambrogi,  E.~Asilar,  T.~Bergauer,  J.~Brandstetter,  E.~Brondolin,  M.~Dragicevic,  J.~Er\"{o},  M.~Flechl,  M.~Friedl,  R.~Fr\"{u}hwirth\cmsAuthorMark{1},  V.M.~Ghete,  J.~Grossmann,  J.~Hrubec,  M.~Jeitler\cmsAuthorMark{1},  A.~K\"{o}nig,  N.~Krammer,  I.~Kr\"{a}tschmer,  D.~Liko,  T.~Madlener,  I.~Mikulec,  E.~Pree,  D.~Rabady,  N.~Rad,  H.~Rohringer,  J.~Schieck\cmsAuthorMark{1},  R.~Sch\"{o}fbeck,  M.~Spanring,  D.~Spitzbart,  W.~Waltenberger,  J.~Wittmann,  C.-E.~Wulz\cmsAuthorMark{1},  M.~Zarucki
% ...
\vskip\cmsinstskip
% ...
1:~Also at Vienna University of Technology,  Vienna,  Austria\\
% ...
\end{sloppypar}
\end{document}